\documentclass[aps,superscriptaddress,amsmath,amssymb,floatfix,showpacs,notitlepage,nofootinbib,preprintnumbers]{revtex4}
\usepackage{graphicx,graphics,setspace,epsfig,color}
\usepackage[letterpaper, dvips,width=7.5in,height=8.5in,includemp=false]{geometry}
\usepackage[vcentermath]{}
\usepackage{epsf}
\usepackage{amscd}
\usepackage{amsmath}
\usepackage{hyperref}

\newcommand{\G}{\mathcal{G}}

\newcommand{\im}{\mathcal{I}{\rm m}}
\newcommand{\re}{\mathcal{R}e}

\newcommand{\kfe}{k_{\rm Fe}}
\newcommand{\kfp}{k_{\rm Fp}}
\newcommand{\kfn}{k_{\rm Fn}}
\newcommand{\kfmu}{k_{{\rm F}\mu}}
\newcommand{\qtf}{q_{\rm TF}}
\newcommand{\qtfe}{q_{\rm TFe}}

\newcommand{\fI}{f_{\rm I}}

\newcommand{\vnI}{V_{\rm nI}}
\newcommand{\vnp}{V_{\rm np}}
\newcommand{\cenp}{\mathcal{C}_{\rm enp}}
\newcommand{\cenI}{\mathcal{C}_{\rm enI}}
\newcommand{\alfa}{\alpha}
\newcommand{\tcn}{T^{\rm n}_{\rm c}}
\newcommand{\tcp}{T^{\rm p}_{\rm c}}

\newcommand{\beq}{\begin{equation}}
\newcommand{\eeq}{\end{equation}}
\newcommand{\bea}{\begin{eqnarray}}
\newcommand{\eea}{\end{eqnarray}}

\begin{document}

\title[title]{Electron-neutron scattering and transport properties of neutron stars}
\author{Bridget Bertoni} 
\email{bbertoni@uw.edu}
\author{Sanjay Reddy}
\email{sareddy@uw.edu}
\author{Ermal Rrapaj}
\email{ermal@uw.edu}
\affiliation{Institute for Nuclear Theory, University of Washington, Seattle, WA}

\preprint{INT-PUB-14-033}
\begin{abstract}
We show that electrons can couple to the neutron excitations in neutron stars and find that this can limit their contribution to the transport properties of dense matter, especially the  shear viscosity.  The coupling between electrons and neutrons is induced by protons in the core, and by ions in the crust. We calculate the effective electron-neutron interaction for the kinematics of relevance to the scattering of degenerate electrons at high density. We use this interaction to calculate the electron thermal conductivity, electrical conductivity, and shear viscosity in the neutron star inner crust, and in the core where we consider both normal and superfluid phases of neutron-rich matter. In some cases, particularly when protons are superconducting and neutrons are in their normal phase, we find that electron-neutron scattering can be more important than the other scattering mechanisms considered previously.       
\end{abstract}



\maketitle
\section{Introduction}
Even before their discovery more than fifty years ago, it was anticipated on theoretical grounds that neutron stars would contain cold and dense matter where quantum effects would manifest on a macroscopic scale \cite{Migdal:1960}. The interpretation of time dependent phenomena observed in x-rays continues to unravel  the connection between such quantum behavior and observable phenomena \cite{PageReddy:2006}. In this context, transport properties of cold, dense matter play a particularly important role because coherence and correlations between particles can have a dramatic impact on the thermal and electrical conductivity, and the shear viscosity of dense matter. For example, shortly  after Bell and Hewish discovered neutron stars  \cite{HewishBell:1968}, Baym, Pethick, and Pines showed that due to extreme degeneracy, matter in the neutron star core would be an excellent electrical conductor, implying that large magnetic fields could be sustained without ohmic dissipation for time scales larger than the  age of the universe \cite{BaymPethickPines:1969b}. In the subsequent decades, several authors, including Flowers and Itoh \cite{FlowersItoh:1976}, and Yakovlev and collaborators \cite{YakovlevUrpin:1980,Shternin:2007,Shternin:2008}, have studied in some detail the electrical conductivity, thermal conductivity, and shear viscosity of neutron star matter. The roles of degeneracy, superfluidity, and superconductivity in the core, and crystalline phases in the lower density regions called the crust, have been studied. Yet, as we argue in this article, an important scattering mechanism between electrons and neutrons, induced by protons in the core and ions in the crust, has been overlooked.  

At the high densities and relatively low temperatures characteristic of neutron stars, a rich phase structure is expected because of strong nuclear and Coulomb interactions, and in some phases, transport processes can be either greatly enhanced or suppressed.  In the outer crust, matter is solid as fully pressure ionized nuclei freeze and form a lattice at low temperature due to strong Coulomb interactions. In the inner crust, where the mass density exceeds $4\times10^{11}$ g cm$^{-3}$, neutrons drip out of nuclei and form a nearly uniform and degenerate Fermi gas alongside the electrons. Due to the strong and attractive s-wave interaction, dripped neutrons in the inner crust are expected to form Cooper pairs and become superfluid at low temperature. The critical temperature for  neutron superfluidity $T^n_c $  is a strong function of density and typical values are in the range $T^n_c\approx 10^8-10^{10}$ K for the densities found in the crust (for a recent review on Cooper pairing of nucleons in neutron stars see \cite{Gezerlis:2014efa}).  

When the mass density  $ \rho \gtrsim 10^{14}$ g cm$^{-3}$, nuclei disappear, possibly through a relatively continuous transition involving non-spherical nuclear shapes, collectively called the pasta phase \cite{PethickRavenhall:1995}. The transition marks the crust-core boundary, with the core containing a liquid phase with degenerate neutrons, protons, and electrons. 
The baryon density in the core is expected to be in the range $ n = n_0/2-4 n_0$, where $n_0=0.16$ nucleons fm$^{-3}$ is the number density inside ordinary nuclei, also called the nuclear saturation density. The typical electron/proton fraction is on the order of a few percent at $n=n_0$, and increases with density. When the electron Fermi energy exceeds the mass of the muon, matter also contains an admixture of muons.  This matter containing neutrons, protons, electrons, and muons in a liquid state may persist up to the highest densities $\approx  10^{15}$ g cm$^{-3}$ at the center of the star in the absence of phase transitions to other exotic forms of matter containing hyperons, kaons, or quarks (for a recent review see \cite{PageReddy:2006}). At the high densities in the core the neutron Fermi momentum is large, and s-wave interactions between neutrons become repulsive.  Superfluid pairing is only possible in p-waves, and p-wave pairing has been found to be fragile and model dependent, with estimates of the critical temperature $\tcn \approx 10^8$ K \cite{SchwenkFriman:2004}. In contrast, because the proton density is small, s-wave interactions remain strongly attractive for protons and s-wave proton superconductivity with a critical temperature $\tcp \approx$ $10^8-10^{10}$ K is expected.

Throughout these different phases of matter inside the neutron star, electrons remain relativistic, degenerate,  and weakly interacting. They consequently play an important role in transport phenomena that shape the thermal, magnetic field, and spin evolution of neutron stars.  In this article we calculate the electron-neutron coupling, which depends on the polarizability of the medium, and show that it is relevant. In earlier work electron-neutron scattering due to the intrinsic magnetic moment of the neutron was considered and found to be unimportant \cite{FlowersItoh:1976}.  Here, in contrast, we find that electron-neutron scattering due to the induced coupling is important for determining the electronic transport properties in the neutron star core.  

We begin with a derivation of the induced interaction in the core and in the inner crust in sections \ref{sec:inducedcore} and  \ref{sec:inducedcrust}, respectively. In section \ref{sec:conductivity} we derive general formulae for the the electron thermal conductivity, electrical conductivity, and shear viscosity. We present our results for these electronic transport properties, and compare them to those obtained in earlier work in section \ref{sec:results}  to highlight situations in which electron-neutron scattering could be important. Our conclusions and some limitations of our study are presented in section \ref{sec:conclusions}. Appendix \ref{app:encore} contains an illustrative derivation of the effective coupling, and in appendix \ref{app:previous} we collect relevant formulae from previous studies which were used to make comparisons. Throughout we use natural units, setting $\hbar = 1$, $c=1$, and $k_{\rm B}=1$, and the electric charge $e=\sqrt{4\pi \alpha}$ where $\alpha=1/137$ is the fine structure constant.  Since the electron Fermi momentum $\kfe \simeq 100 ~{\rm MeV} \gg m_e$  for typical densities encountered in the neutron star inner crust and core, throughout we treat the electrons as ultra-relativistic particles with velocity $v_e=c=1$.


\section{Induced electron-neutron interaction in the core} 
\label{sec:inducedcore}

In free space, and at low momenta, the electron-neutron interaction is weak as it arises due to the small neutron magnetic moment. In contrast, in the dense plasma inside neutron stars, electrons can couple to neutrons due to an interaction induced by the polarizability of the charged protons.  This can be understood intuitively by noting that the presence of the neutron in the medium will disturb its immediate vicinity, and affect in particular the proton density distribution. This will create either a positively or negatively charged cloud around the neutrons depending on whether the neutron-proton interaction is attractive or repulsive. At low density since the neutron-proton interaction is attractive, the neutron will acquire a net positive charge while at the high density where the interaction can be repulsive, the charge cloud surrounding the neutron will be negative. 

\begin{figure}[htbp] 
   \centering
   \includegraphics[width=2in]{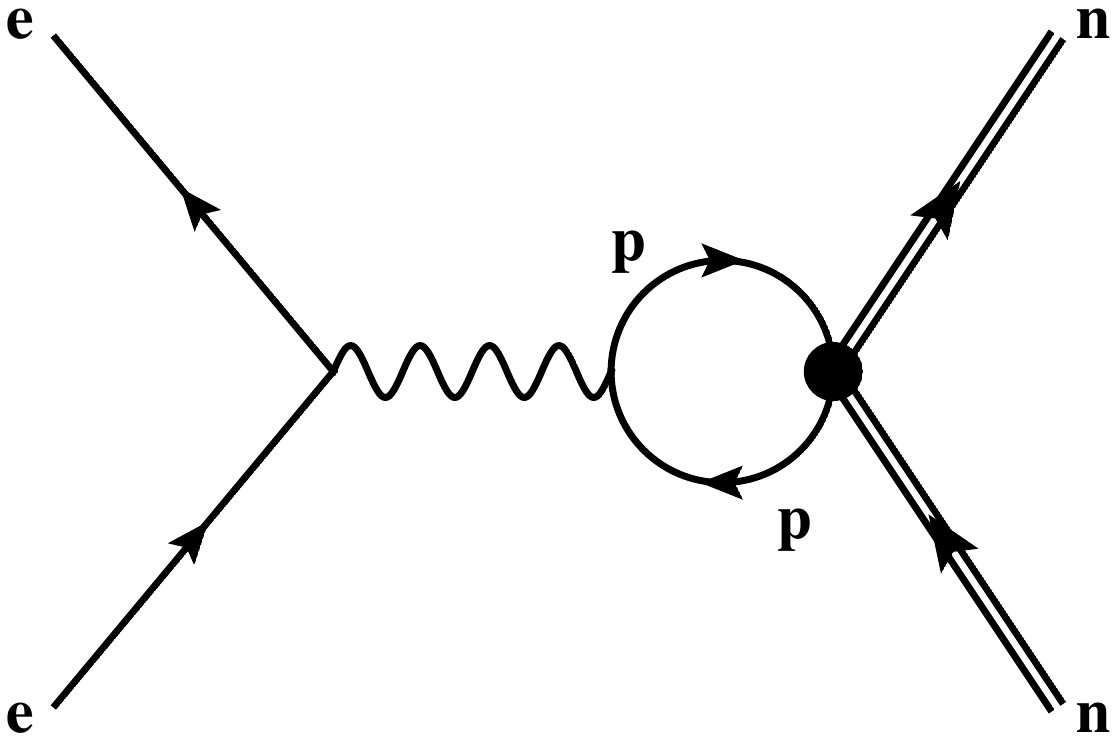} 
   \caption{Effective interaction between electrons and neutrons induced by protons in the medium. The wavy line represents the  plasmon.}
   \label{fig:enp_diagram}
\end{figure}

The effective coupling between neutrons and electrons is mediated by the in-medium photon (the plasmon) and can be derived using standard techniques in quantum field theory (for more details see appendix \ref{app:encore}).  An effective Lagrangian for the electron-neutron coupling can be derived in analogy with the plasmon-neutrino coupling described in Refs.~\cite{Braaten:1993,AltherrSalati:1994}. The Feynman diagram in Fig.~\ref{fig:enp_diagram} shows the the exchange of a plasmon (wavy-line) which couples electrons to protons. 
The protons in turn couple to neutrons by the short-range strong interaction depicted by a filled circle.  From the diagram it follows that the plasmon-neutron coupling can be described by the effective Lagrangian
\beq
{\cal L}_{\gamma \rm -n} = -\sqrt{4\pi \alfa}~V_{\rm np} ~\bar{n} \gamma_\mu n ~ \Pi_{\rm p}^{\mu \nu} A_\nu \,,
\label{eq:Lphn}
\eeq
where $n$ and $A_\nu$ are the neutron and plasmon fields, $V_{\rm np}$ is the short-range nuclear potential, and $\Pi_{\rm p}^{\mu \nu}$ is the proton polarization correction to the photon in the plasma which can be decomposed into longitudinal and transverse components and is given by 
\beq
\Pi_p^{\mu \nu}(\omega,q)=\Pi^L_p(\omega,q)\left(1,\frac{\omega}{q}\hat{q}\right)^\mu \left(1,\frac{\omega}{q}\hat{q}\right)^\nu + \Pi^T_p(\omega,q)~g^{\mu i}(\delta^{ij}-\hat{q}^i\hat{q}^j)g^{j \nu}
\,,
\eeq    
where $i,j$ are spatial indices and the four-momentum $q^\mu=(\omega,q\hat{q})$ \cite{Kapusta}. $\Pi^L_p$ and $\Pi^T_p$ are complex functions in general but we neglect the imaginary part in defining the plasmon-neutron coupling for the following reasons. When the protons are in the normal phase, the imaginary part corresponds to real proton excitations and leads to Landau damping of the plasmon. Its magnitude is small (proportional to $ m^2_p~\omega /q$ where $m_p$ is proton mass) and vanishes in the static limit $\omega \rightarrow 0 $ \cite{FetterWalecka:2003}. Instead if protons are superconducting, the imaginary part is zero for $\omega < 2 \Delta_p$ where $\Delta_p$ is the energy gap in the proton spectrum \cite{Schrieffer:1964}. The real part of the longitudinal polarization function is denoted as $\chi_p(\omega,q)$ and is given by 
\beq
\chi_p(\omega,q)= \re~\Pi^L_p(\omega,q) =\re ~\int dt ~e^{i\omega t}\int d {\bf r} ~e^{-i {\bf q \cdot r}}~\langle n_p({\bf r},t) n_p(0,0) \rangle \,,
\label{eq:chi}
\eeq
where $n_p=\bar{p}\gamma_0 p=p^\dagger p$ is the proton density operator. The real part of the transverse polarization function is related to the velocity-velocity correlation function of the proton fluid.  Note that $q_\mu \Pi^{\mu\nu}=0$ so that this effective electron-neutron coupling is manifestly  gauge invariant. 

Since the proton fraction in neutron stars is typically only a few percent, the proton Fermi momentum is small and protons can be treated in the non-relativistic limit. The neutrons are also only mildly relativistic in the vicinity of nuclear saturation density with a velocity  $v_{\rm Fn} = \kfn/m_n \simeq 1/3$. This, together with the fact that scattering kinematics is restricted to the region $\omega < q$, implies that it is reasonable to neglect the spatial components of the currents in Eq.~\ref{eq:Lphn}. Retaining only the density-density component, the effective interaction is described by the Lagrangian  
\beq
{\cal L}_{\rm e-n} = - \bar{e}\gamma_0 e~ \mathcal{U}_{\rm enp}(\omega,q)  ~\bar{n} \gamma_0 n \,,
\label{eq:Len0}
\eeq
where $e$ is the electron field and 
\beq
\mathcal{U} _{\rm enp}(\omega,q)
=\frac{-4\pi \alfa~\cenp(\omega,q) }{q^2+\qtf^2} \,;~~ \cenp(\omega,q) = V_{\rm np}(q) \chi_p(\omega, q) \,,
\label{eq:cepn}
\eeq 
is the induced interaction, and the Thomas-Fermi momentum $\qtf$ includes the screening of electric charge due to protons, electrons, and muons, and is defined by 
\beq
\qtf^2=4\pi \alfa \sum_{\text{ i=e,p,}\mu} \frac{\partial n_i}{\partial \mu_i} \approx \frac{4\alfa}{\pi} \left( m_p\kfp + \kfe^2 +k_{\text{F$\mu$}}\sqrt{m_{\mu}^2+k_{\text{F$\mu$}}^2} \right) \,.
\eeq 

We now turn to discuss $\chi_p$ and $\vnp$, both of which are needed to calculate the effective electron-neutron coupling $\cenp(q)$, which characterizes the strength of the electron-neutron interaction relative to the Coulomb interaction between electrons and protons. First, we note that due to the strong degeneracy of electrons at low temperature, the energy transfer $\omega \approx T $ is small compared to the momentum transfer $q$ and other relevant energy scales associated with the dense medium. For this reason, the effective coupling  can be calculated in the static limit. In this limit, the susceptibility $\chi_p(q)\equiv \chi_p(\omega = 0 ,q)$ is well known from non-relativistic many-body theory for a non-interacting Fermi gas and is given by   
\beq 
\chi_p(q)= \re~ \Pi_p^0(\omega=0,q)= -\frac{m_p k_{\rm Fp}}{2\pi^2}~\left(1+\left(\frac{1-y^2}{2y}\right)\log{\left|\frac{1+y}{1-y}\right|}\right)\,,
\label{eq:chi_p}
\eeq  
where the one-loop polarization function $\Pi_p^0$ is defined in Eq.~\ref{eqn:pi_0} and  $y=q/2 k_{\rm Fp}$ \cite{FetterWalecka:2003}. In a strongly interacting system, higher order corrections to the one-loop polarization function can become relevant but are known not change the qualitative behavior. We estimate the size of these corrections by noting that in the long-wavelength limit $\chi_p(q\rightarrow 0)=-\partial n_p/\partial \mu_p$ where $n_p$ and $\mu_p$ are the proton number density and chemical potential, respectively. Using microscopic calculations of the dense matter equation of state (EoS) reported in Refs. \cite{Akmal:1998cf} and \cite{Gandolfi:2009nq} we have calculated this derivative to find corrections in the range $20\% -50\%$ in the vicinity of  $n=n_0$, and an enhancement by a factor of two at the highest densities ($n \simeq 0.48$ fm$^{-3}$) encountered in the core. In contrast, corrections to $\chi_p(q)$ due to proton superconductivity are small in the static limit for typical values of the superconducting gap $\Delta_p \simeq 1 ~{\rm MeV} \ll \mu_p$ and can be safely neglected \cite{Schrieffer:1964}.

The potential $\vnp(q)$ describes the interaction between neutrons and protons in the medium and is in general a complicated function of density and momentum. However, later in section \ref{sec:conductivity} we shall find that typical momentum transfer involved in electron collisions is in the range of a few times $\qtf$, and this justifies a low momentum expansion of the form 
\beq
\vnp(q) = \vnp^{(0)} + \vnp^{(2)}~\frac{q^2}{\kfe^2}\,. 
\label{eq:vexp2}
\eeq
The effective interaction at zero momentum exchange can be extracted from the EoS of dense matter through the relation 
\beq 
\vnp^{(0)} =\vnp(q\rightarrow 0) = \frac{\partial^2{{\cal E}}}{{\partial n_p \partial n_n}} \,, 
\label{eqn:vnpq0}
\eeq  
where ${\cal E}(n_n,n_p)$ is the energy density of the liquid of neutrons and protons with density $n_n$ and $n_p$, respectively.  We use EoS models described in Refs.~\cite{Akmal:1998cf} and \cite{Gandolfi:2009nq}, which are based on non-perturbative calculations using realistic two and three nucleon interactions, to calculate ${\cal E}(n_n,n_p)$ and the second derivative $\partial^2 {\cal E}(n_n,n_p)  / \partial n_n \partial n_p$.  Numerical differences between them can be viewed as a rough error estimate and for this reason we shall present results for both EoSs. The second term in the expansion $\vnp^{(2)}$ cannot be derived from the EoS, but it is related to the $L=1$ Fermi liquid parameters. Since we are unaware of a microscopic calculation of these parameters in neutron-rich matter, we have opted to use a range $\vnp^{(2)}=5\times 10^{-6}-5\times 10^{-5}$ MeV$^{-2}$ as suggested by calculations in symmetric nuclear matter from Ref.~\cite{Holt:2011yj}. 

\begin{figure}[h] 
   \centering
 \includegraphics[width=3.7in]{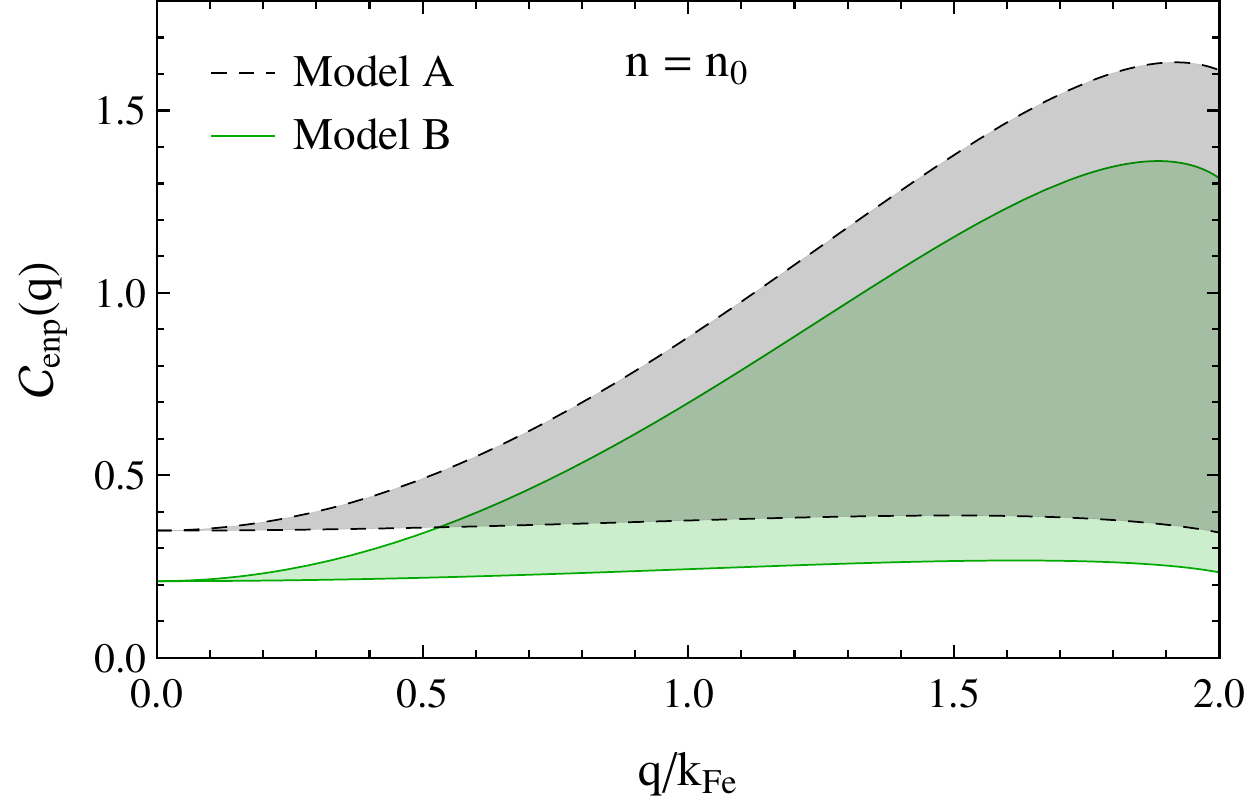} 
 \includegraphics[width=3.6in]{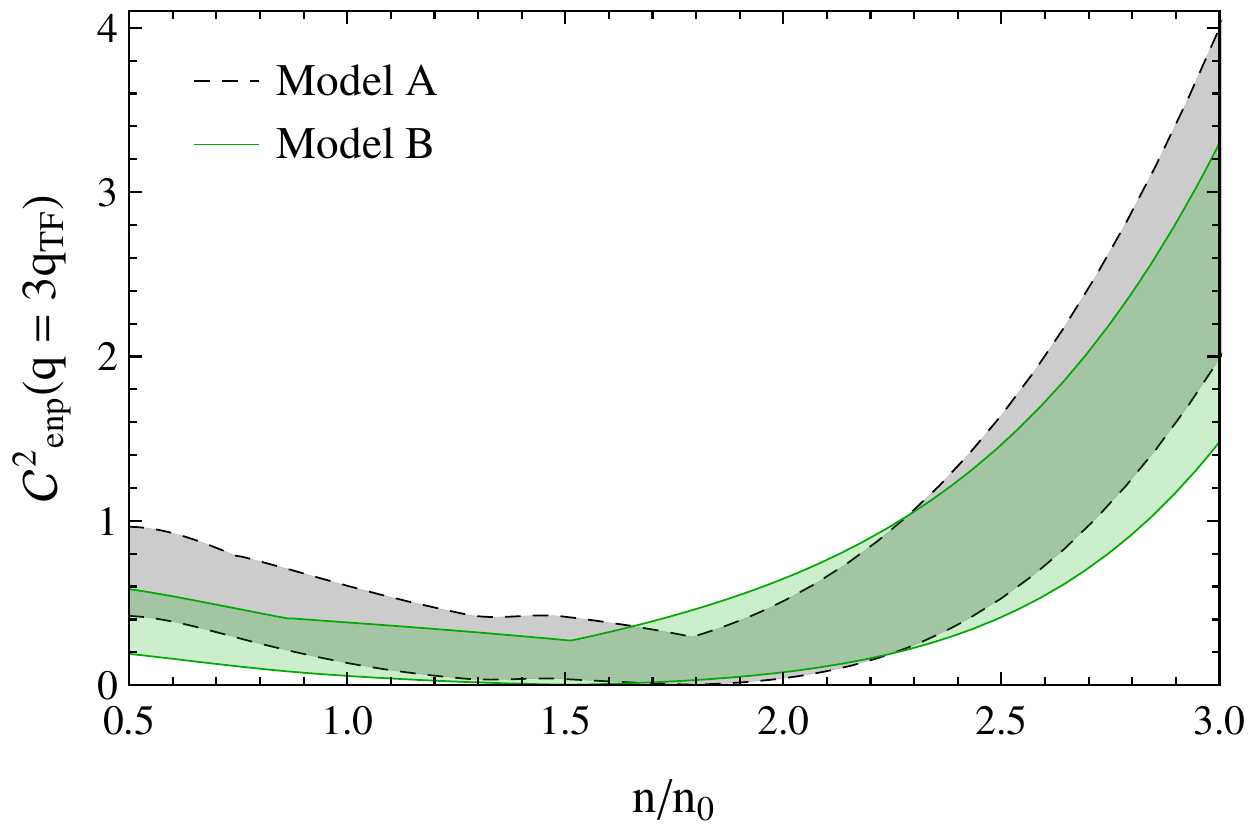} 
   \caption{Left panel: The strength of the induced electron-neutron interaction defined in Eq.~\ref{eq:cepn} as a function of $q$ at nuclear saturation density $n_0=0.16$ fm$^{-3}$. Right panel: $\cenp^2(q=3\qtf)$ as a function of density.}
\label{fig:cenp}
\end{figure}
The momentum and density dependence of the strength of the induced coupling $\cenp$ is shown in Fig.~\ref{fig:cenp} for the two models defined as follows. 
In model A, we use the EoS from Ref.~\cite{Akmal:1998cf} to obtain $\vnp^{(0)}$ and the band is obtained by varying $\vnp^{(2)}$ over the range mentioned above. In model B,   we use the EoS from Ref.~\cite{Gandolfi:2009nq} to obtain $\vnp^{(0)}$ and the band is obtained by varying $\vnp^{(2)}$ over the same range.      

\section{Electron-neutron interaction in the crust} 
\label{sec:inducedcrust}
In the inner crust, free protons are absent as they are all bound into neutron-rich ions.  Here, the interaction between electrons and neutrons is induced by these ions and arises due to the nuclear interaction between the ions and the unbound neutrons. The calculation of the induced interaction proceeds along the lines described earlier, but with ions in the crust playing the role of protons in the core. We find, as before, that the induced interaction can be written in the form 
\beq
\mathcal{U} _{\rm enI}(0,q)
= \frac{-4\pi \alfa~ \cenI(q) }{q^2+\qtfe^2}; \quad \cenI(q) = V_{\rm nI}(q)~Z~ \chi_I(q)\,,   
\label{eq:enI_q}
\eeq
where $q_{\rm TFe}=\sqrt{4\alfa/\pi} ~k_{Fe}$ is the Thomas-Fermi screening momentum of the electron gas, $V_{\rm nI}(q)$ is the short-range neutron-ion interaction, $Z$ is the charge of the ion, 
and 
\beq
\chi_I(q)= \re~\int dt\int d {\bf r} ~e^{-i {\bf q \cdot r}}~\langle n_I({\bf r},t) n_I(0,0) \rangle \,,
\eeq
is the ion static density-density correlation function where $n_I(r,t)$ is the ion density operator. The strength and momentum dependence of the interaction  $V_{\rm nI}(q)$ will in general depend on the nuclear structure of the  extreme neutron-rich nuclei present in the crust and has not been studied in any detail. Since the typical momentum transfer $q \simeq$ a few times $\qtfe$, and $\qtfe$ is small compared to $\kfe$, a momentum expansion of the form in Eq.~\ref{eq:vexp2} can be justified. Again, the leading order term, which is independent of momentum, can be calculated from the EoS through the relation $V_{\rm nI}^{(0)} = \partial^2 {\cal E}  / \partial n_n \partial n_I $  where $\cal E$ is the energy density. We use the data from \cite{kobyakov2013dynamics} for the composition and EoS of the crust to calculate $V_{\rm nI}^{(0)}$. However, as we are unaware of any calculations of $V_{\rm nI}^{(2)}$ that we can make use of here, and its calculation is beyond the scope of this study, we set $V_{\rm nI}^{(2)}=0$ to obtain a conservative estimate of $\cenI(q)$. 

$\chi_I(q)$ in the crust differs from $\chi_p(q)$ in the core because Coulomb correlations between ions are strong enough to result in crystalline structure at low temperature. To compute $\chi_I(q)$ in the crystalline state we can use the fact that the ion dynamic form factor $S_I (\omega,q)$ is dominated by phonons, and we can combine this with the fluctuation-dissipation theorem and the Kramers-Kronig relation to find that 
\beq
\chi_I(q)= -\mathcal{P}~\int_{-\infty}^{\infty}\frac{d\omega'}{2\pi}\frac{(1-e^{-\omega'/T})S_I(\omega',q)}{\omega'} \,.
\eeq 
In the one-phonon approximation, 
\beq
S_I(\omega,q) = \frac{n_I}{M_I}~\sum_{\lambda, \bf K, \bf k}\frac{\pi(\epsilon_\lambda({\bf k})\cdot {\bf q})^2}{\omega_{k\lambda}}\bigg[\frac{\delta(\omega-\omega_{k\lambda})\delta_{\bf q, \bf k + \bf K}}{1-e^{-\omega/T}}+\frac{\delta(\omega+\omega_{k\lambda})\delta_{\bf q, \bf k + \bf K}}{e^{-\omega/T}-1}\bigg]\,,
\eeq
where the sum is over the reciprocal lattice vector ${\bf K}$, the three polarization states $\lambda$ of the phonon spectrum $(\omega_{k\lambda} = v_{\lambda}k)$ with velocity $v_\lambda$, and the vector $ {\bf k}$  is restricted to be in the first Brillouin zone. $n_I$ is the ion density and $M_I$ is the ion mass. 

Approximating the lattice of nuclei as simple cubic with lattice spacing $a = n_I^{-1/3}$, we find that
\beq \label{eq:chi_I}
\begin{split}
\chi_I(q) &= -\frac{n_I}{M_I}~\sum_{\lambda, \bf K, \bf k}\frac{(\epsilon_\lambda({\bf k})\cdot {\bf q})^2}{\omega_{k\lambda}^2}\delta_{\bf q, \bf k + \bf K} \\
&= -\frac{n_I}{M_I}~\sum_{\lambda, \bf K}\frac{(\epsilon_\lambda({\bf q - \bf K})\cdot {\bf q})^2}{\omega_{|\bf q - \bf K|\lambda}^2} \prod_{i=x,y,z} \theta \left(\frac{\pi}{a} - |{\bf q}_i - {\bf K}_i|\right) \,,
\end{split}
\eeq
where ${\bf q}_i$ and ${\bf K}_i$ are the i$^{\rm th}$ components of the respective vectors.  The corresponding ion density, mass, and charge, and the two transverse and one longitudinal phonon velocities can be calculated following \cite{Chamel:2013}.  The velocity of the longitudinal mode of lattice vibrations is $ v_l\approx \omega_P/q_{\rm TFe}$ where $\omega_P=\sqrt{4\pi \alfa Z^2 n_I/M_I}$ is the ion plasma frequency.  The velocity of the two transverse modes is $v_t \approx 0.4 ~ \omega_P/q_D$ where $q_D = (6\pi^2n_I)^{1/3}$ is the Debye wavenumber.  In general, both the longitudinal and the two transverse modes contribute to the ion response.  However, for $q< \pi/a$, Eq.~\ref{eq:chi_I} simplifies since only one term of the sum, which corresponds to ${\bf K}=0$ with longitudinal polarization, survives and we find that 
\beq
\chi_I(q<\pi/a)=-\frac{n_I}{M_I v_l^2} \,.
\label{eq:chiI0}
\eeq
For $q \gg \frac{\pi}{a}$ we found that converting the sum over $\bf K$ to an integral and confining $|\bf q - \bf K|$ to be within a sphere of diameter $\pi/a$ instead of a cube with side length $\pi/a$ gave an analytic, conservative lower limit on $\chi_I(q)$ given by
\beq 
\chi_I(q \gg \pi/a) =  -\frac{n_I} {M_I v_t^2} \left(\frac{4 \pi}{3}\right)^{1/3} \left(1 + \frac{v_t^2}{2v_l^2}\right)  \left(\frac{q}{q_D}\right)^2 \,.
\label{eq:chiI2}
\eeq
This result combined with the fact that $v_t \ll v_l$ implies that $\chi_I(q > \pi/a)$ increases rapidly with momentum. 
\begin{figure}[h] 
\includegraphics[width=3.7in]{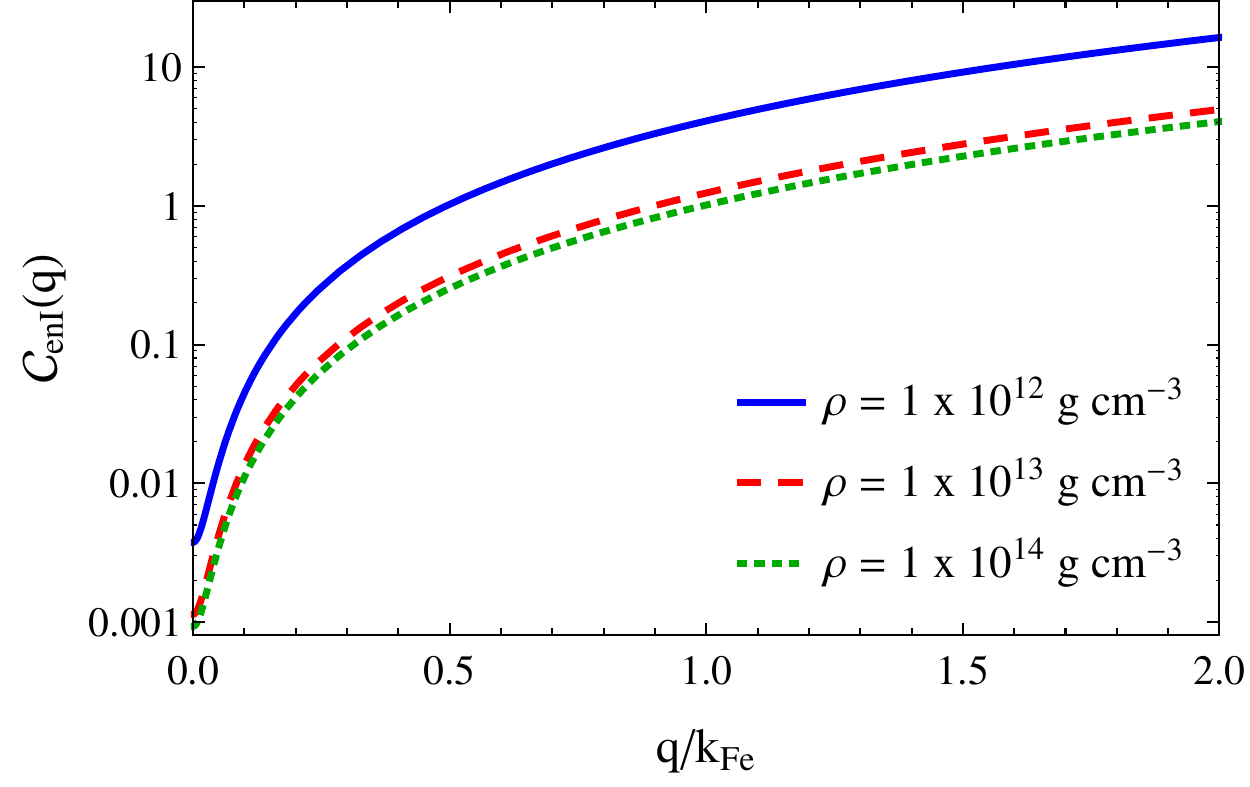} 
\includegraphics[width=3.6in]{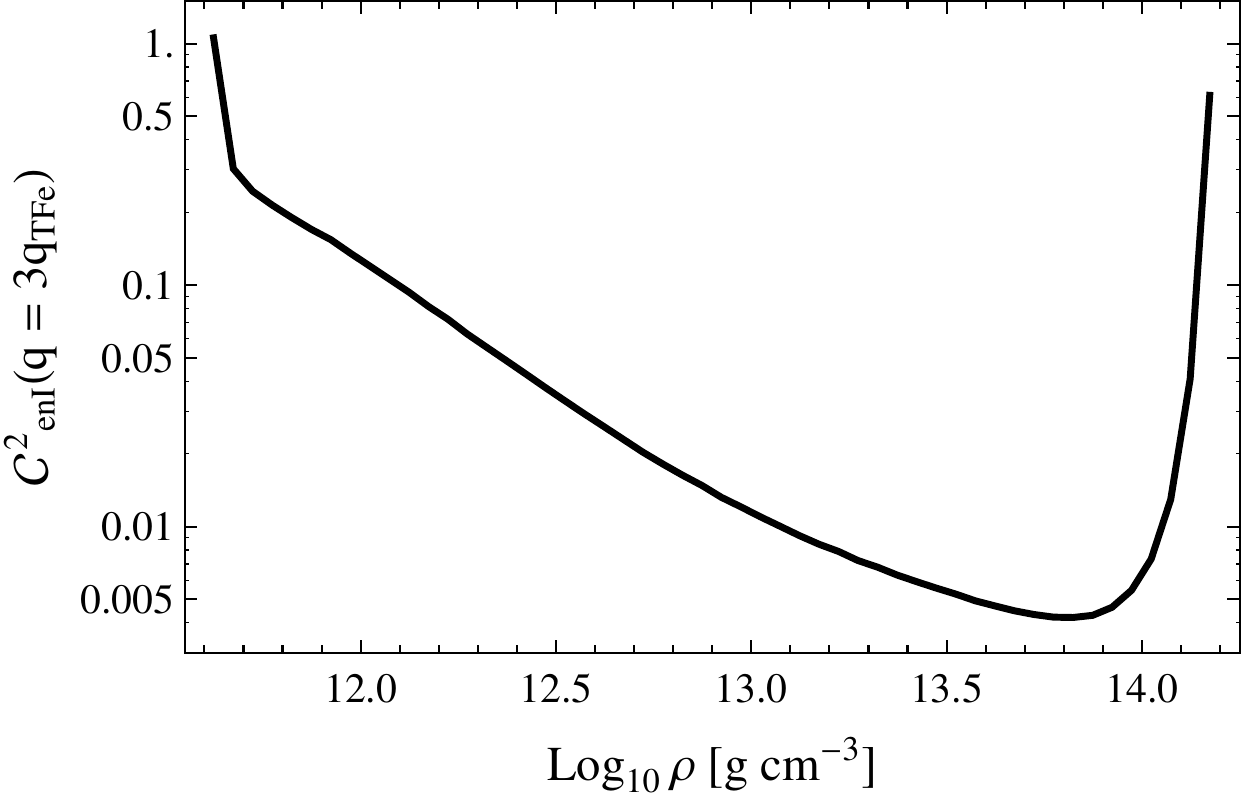} 
\caption{Left panel: The strength of the induced electron-neutron interaction in the crust as a function of $q$ at various densities in the crust. Right panel: $\cenI^2(q=3\qtfe)$ as a function of density.}
\label{fig:ceniq}
\end{figure}

The momentum dependence of $\cenI (q)$ is plotted in the left panel of Fig.~\ref{fig:ceniq} for typical densities encountered in the crust and arises solely due to large momentum dependence $\chi_I(q)$.  The right panel of Fig.~\ref{fig:ceniq} shows the density dependence of $\cenI$ evaluated at a momentum $q=3\qtfe$ which is expected to be a typical scale. Here, the large density dependence is entirely due the EoS of the crust from Ref.~\cite{kobyakov2013dynamics}.    

\begin{figure}[htbp] 
   \includegraphics[width=2in]{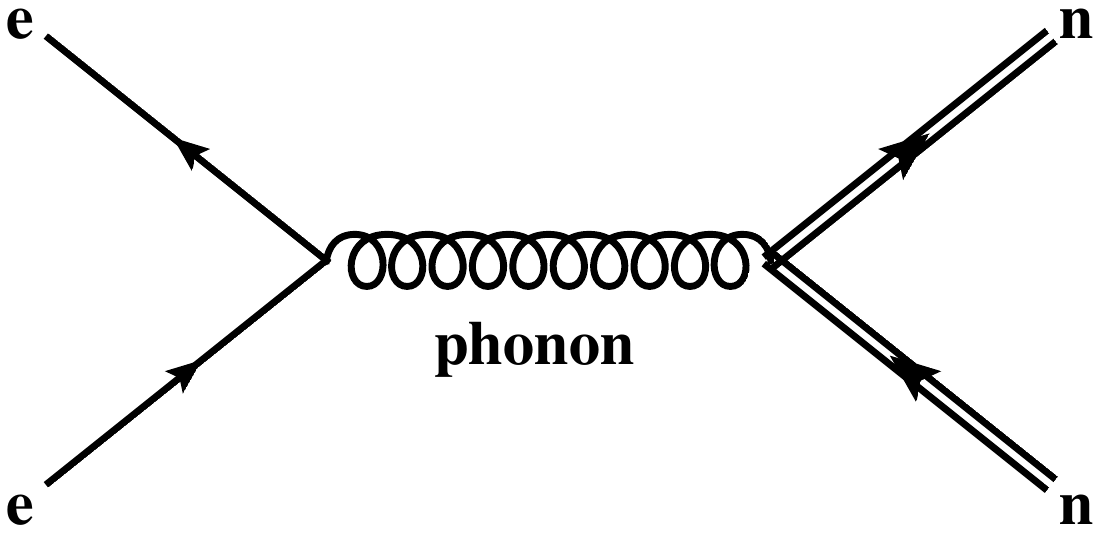} 
   \caption{Effective interaction between electrons and neutrons induced by ion density fluctuations represented by a phonon.}
   \label{fig:en_diagram}
\end{figure}

Before discussing the implications of the electron-neutron coupling we present an alternate, perhaps simpler and more instructive, derivation of the electron-neutron interaction which follows from noting that at low momentum, the interaction is mediated by one-phonon exchange. The process is shown by the Feynman diagram in Fig.~\ref{fig:en_diagram} where the single and double fermion lines represent electrons and neutrons, respectively and the exchange particle is the longitudinal phonon of the ion lattice. 

The electron-phonon coupling is well known, and at small momentum it is described by the Lagrangian 
\beq
\mathcal{L} _{\rm e-ph}
= \frac{4\pi Z \alfa~\fI}{q^2+\qtfe^2}~e^\dagger e~\nabla_i \xi_i \,, 
\label{eq:epn1}
\eeq 
where $\xi_i$ is the canonically normalized phonon field which is related to ion density fluctuations, $\delta n_I = -\fI \nabla_i \xi_i$, and $\fI=\sqrt{n_I/m_I}$ \cite{FetterWalecka:2003}. Density fluctuations of the ion lattice also couple to the neutrons due to the short-range neutron-ion potential  $\vnI$. The coupling between low energy neutrons and lattice phonons is described by the Lagrangian   
\beq
\mathcal{L} _{\rm n-ph}= -\vnI ~\fI~n^\dagger n~\nabla_i \xi_i \, .
\label{eq:epn2}
\eeq

The electron-neutron interaction follows from Fig.~\ref{fig:en_diagram}, where the phonon-electron and phonon-neutron vertices are given by Eq.~\ref{eq:epn1} and Eq.~\ref{eq:epn2}, respectively. Using the longitudinal phonon propagator ${\cal D}(q) = (\omega^2-\omega^2(q))^{-1}$ where the phonon dispersion relation is $\omega^2(q) = v_l^2q^2$ and $v_l$ is the velocity of the longitudinal mode, we find that the induced interaction is 
\beq 
\mathcal{U} _{\rm enI}(\omega,q)=\left(\frac{-4\pi Z\alfa\vnI\fI^2}{q^2+\qtfe^2}\right)~\frac{q^2}{(\omega^2-v_l^2 q^2)} \,. 
\label{eq:uenI}
\eeq
In the static limit where $\omega \ll v_l q$ this simplifies to 
\beq
\mathcal{U} _{\rm enI}(0,q)=\frac{4\pi \alfa Z \vnI \fI^2}{ (q^2+\qtfe^2)v_l^2} \,, 
\eeq
which coincides with the result obtained earlier in Eq.~\ref{eq:enI_q} because $\cenI(q)= -\vnI Z  f^2_I/v^2_l$ for small $q<\pi/a$.


\section{Electron conductivities and shear viscosity}  
\label{sec:conductivity}
We now turn to the calculation of the electron thermal conductivity, electrical conductivity, and shear viscosity due to electron-neutron scattering. Both in the crust and in the core, the thermal and electrical conductivity, and the shear viscosity are given by 
\beq
\kappa_{\rm en}  = \frac{\kfe^2~T}{9}~\langle \lambda_{en} \rangle_{\kappa}\,, \quad
\sigma_{\rm en}  = \frac{\kfe^2~\alfa}{3\pi^2}~\langle \lambda_{en} \rangle_{\sigma}\,, \quad
\eta_{\rm en} = \frac{k_{\rm Fe}^4}{15\pi^2}~\langle \lambda_{en} \rangle_{\eta}\,,
\eeq       
respectively, and these expressions are written in a form familiar from kinetic theory. The relevant transport mean free paths are obtained as simple variational solutions to the Boltzmann equation for degenerate and relativistic electrons \cite{FlowersItoh:1976}. For electron-neutron scattering described by an effective interaction of the form in Eq.~\ref{eq:cepn}, the mean free paths are given by    
\bea
\langle \lambda_{en} \rangle_\kappa^{-1} &=& \frac{1}{4\pi \kfe^2}\int_0^{2k_{\rm Fe}} dq ~q^3\frac{(4\pi \alfa)^2\mathcal{C}_{\rm en}^2(q)}{(q^2+\qtf^2)^2} \left(1- \frac{q^2}{4 \kfe^2}\right)~I_{\kappa}(q) \,, 
\label{eqn:Lambda_e_kappa} \\
\langle \lambda_{en} \rangle_\sigma^{-1} &=& \frac{1}{ 4 \pi \kfe^2}\int_0^{2k_{\rm Fe}} dq ~q^3\frac{(4\pi \alfa)^2\mathcal{C}_{\rm en}^2(q)}{(q^2+\qtf^2)^2} \left(1- \frac{q^2}{4 \kfe^2}\right) ~I_{\sigma}(q) \,, 
\label{eqn:Lambda_e_sigma}\\
\langle \lambda_{en} \rangle_\eta^{-1} &=& \frac{1}{4\pi \kfe^2}\int_0^{2k_{\rm Fe}} dq ~q^3\frac{(4\pi \alpha)^2\mathcal{C}_{\rm en}^2(q)}{(q^2+q_{\rm TF}^2)^2} \left(1- \frac{q^2}{4 \kfe^2}\right)~I_{\eta}(q) \,,
\label{eqn:Lambda_e}
\eea
respectively. Here, the strength of the induced coupling in the general case is defined as $\mathcal{C}_{\rm en}$ which stands for $\cenp$ in the core and $\cenI$ in the crust, and 
\beq
I_{\kappa/\sigma/\eta}(q)= \int_{-\infty}^{\infty} \frac{d\omega}{2\pi}~ \frac{\beta \omega}{e^{\beta \omega}-1}S_n(\omega,q) ~g_{\kappa/\sigma/\eta}  \,,
\label{eq:Is}
\eeq
where 
\begin{eqnarray}
&g_\kappa = 1+\left(\frac{\beta \omega}{\pi}\right)^2 \left(3 \frac{\kfe^2}{q^2}- \frac{1}{2}\right) \,, \quad 
g_\sigma=1 \,, \quad g_{\eta} = 3\left(1-\frac{q^2}{4k_{\rm Fe}^2}\right) \,,
\label{eqn:gkappa}
\end{eqnarray}
$\beta = 1/T$, and $S_n(\omega,q)$ is the dynamical structure function describing the spectrum of neutron density fluctuations.  The factor of $(1-q^2/(4\kfe^2))$ in the mean free paths arises because for ultra-relativistic particles helicity is conserved and this suppresses the back-scattering of electrons. The dynamical structure function of the degenerate neutron gas in the normal and superfluid phases differs greatly. In the normal phase, the response is dominated by particle-hole excitations at the Fermi surface. In the superfluid phase, because particle-hole excitations are suppressed by the gap, the response is dominated by a collective mode (the Goldstone boson associated with the spontaneous breaking of the $U(1)$ symmetry in the superfluid state) called the superfluid phonon. 

We will begin by first considering the dynamical  structure function in the normal phase which is given by  
\beq
S_n(\omega,q) = \frac{-2~\im~\Pi^0_n(\omega,q)}{1-\exp{(-\beta \omega)}} ~\,, 
\eeq
where $\Pi^0_n(\omega,q)$ is the neutron density-density correlation function and is defined in appendix \ref{app:encore}. In the special case when interactions can be neglected the dynamical structure function is given by 
 \beq
S^{\rm FG}_n(\omega,q) = 2 \int\frac{d^3p_1}{(2\pi)^3}\int\frac{d^3p_2}{(2\pi)^3}(2\pi)^4\delta^4(q^\mu +p_1^\mu-p_2^\mu)f(E_1)(1-f(E_2)) \,,
\eeq
where $p_1$ and $p_2$ are initial and final neutron momenta, $E_1$ and $E_2$ are initial and final neutron energies, $q^{\mu} = (\omega,\vec{q})$, and $f(E)$ is the Fermi-Dirac distribution function of the non-interacting neutron gas. Further, when $\omega \ll \mu_n$ and $q \ll k_{\rm Fn}$  interactions can be incorporated within the framework of Fermi liquid theory. In this case 
\beq 
S_n(\omega,q) \approx \frac{(1-e^{-\beta\omega})^{-1}}{(1+F_0)^2}~ \frac{\left(m^*_{n}\right)^2~\omega}{\pi q}~\Theta(q v_{\rm Fn} - |\omega|)\,,
\label{eq:normal_s}
\eeq
where $v_{\rm Fn}=\kfn/m_n^*$ is the neutron Fermi velocity, and the effective mass $m_n^* \simeq m_n$ and the factor $F_0 \approx -0.5~-~1$ are Fermi liquid parameters for the conditions encountered in neutron stars  \cite{IwamotoPethick:1982}. In the following  we shall neglect these Fermi liquid corrections by setting $F_0=0$ and $m_n^* = m_n$ in Eq.~\ref{eq:normal_s}.  Using this result we can explicitly write the integral defined in Eq.~\ref{eq:Is} as  
\beq
I_{\kappa/\sigma/\eta}(q)= \frac{m_n^2}{\pi^2 \beta q} \int_{0}^{q v_{\rm Fn}} d\omega~ \frac{(\beta \omega)^2}{(e^{\beta \omega}-1)(1-e^{-\beta \omega})}  ~g_{\kappa/\sigma/\eta} \,. \\
\eeq
In the low temperature limit when $T \ll \kfe v_{\rm Fn} $ the integral over $\omega$ is performed by setting the upper limit to $\infty$ and the integral becomes independent of the neutron Fermi momentum. We find that 
\bea
I_{\kappa}(q)= \frac{1}{5}\frac{m_n^2}{ \beta^2 q} ~\left(1+\frac{4 \kfe^2}{q^2} \right)\,, \quad
I_{\sigma}(q)= \frac{1}{3}\frac{m_n^2}{ \beta^2 q}\,, \quad
I_{\eta}(q) = \frac{m_n^2}{\beta^2 q}~\left(1-\frac{q^2}{4k_{\rm Fe}^2}\right) \,.
\label{eq:Ins} 
\eea
It is interesting to note the difference between $I_{\kappa}(q)$ and $I_{\sigma}(q)$ in the above. This difference arises solely due to the inelasticity of electron-neutron collisions since the energy transfer $\omega \simeq q v_{\rm Fn} \simeq T$ is favored and this implies that the Wiedemann-Franz law will be violated for electron-neutron scattering. 

\begin{figure}[h] 
   \centering
    \includegraphics[width=3.7in]{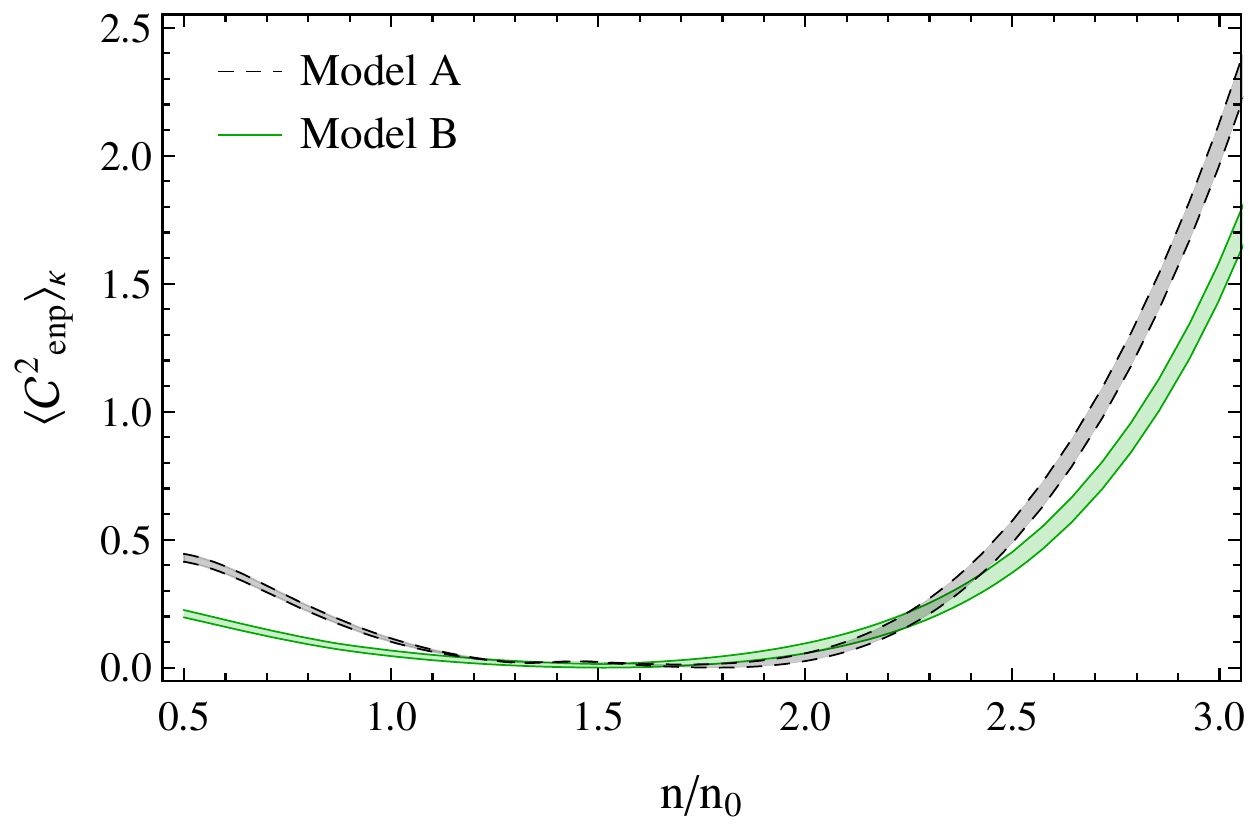} 
   \includegraphics[width=3.7in]{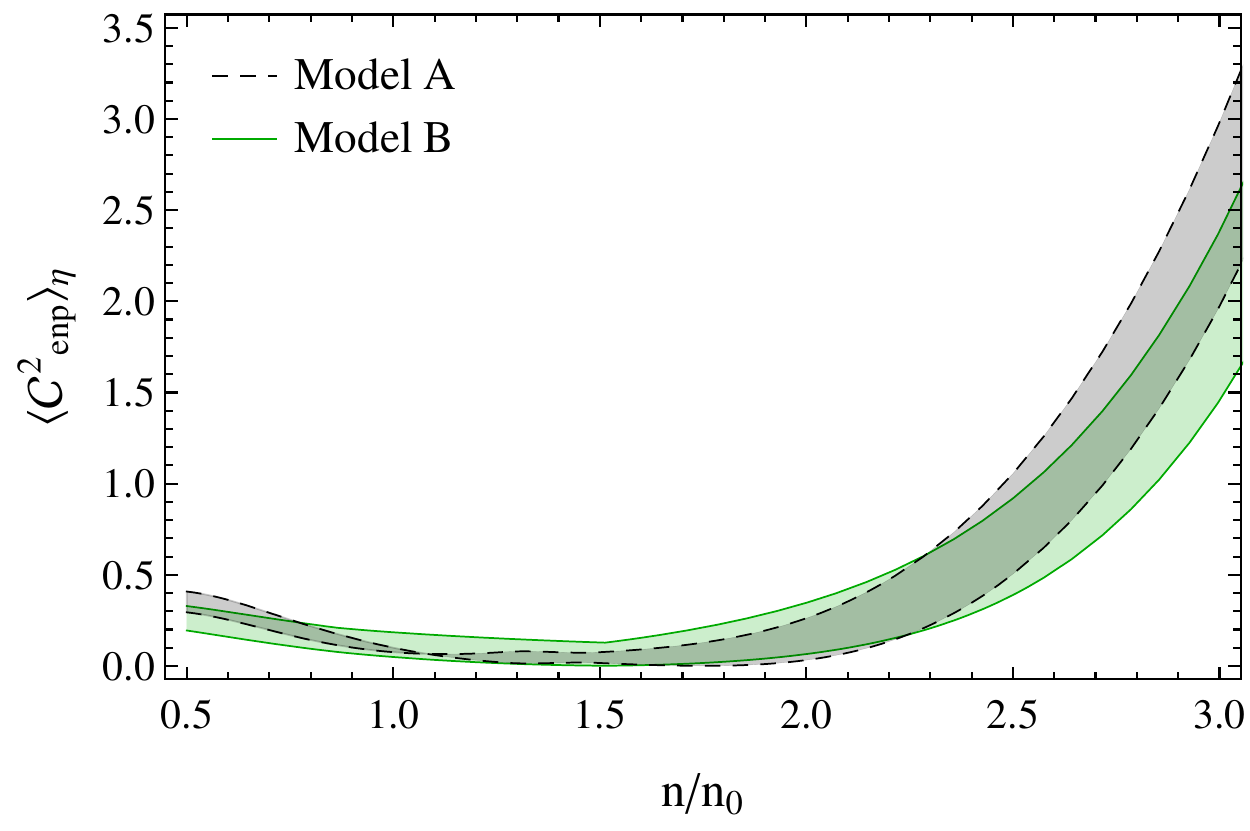} 
   \caption{The momentum averaged couplings $ \langle \mathcal{C}_{\rm en}^2\rangle_{\kappa}$ (left panel) and $\langle \mathcal{C}_{\rm enp}^2\rangle_{\eta}$ (right panel) for densities of relevance in the core. It is assumed that neutrons are in the normal phase and Eq.~\ref{eq:ceq_av} is used to obtain the averages. Although it is not shown we remark that $\langle \mathcal{C}_{\rm enp}^2\rangle_{\sigma} \approx \langle \mathcal{C}_{\rm enp}^2\rangle_{\eta} $. }
   \label{fig:cenp_core}
\end{figure}

It is convenient to define the following momentum averaged effective couplings 
\beq
 \langle \mathcal{C}_{\rm en}^2\rangle_{\kappa/\sigma/\eta} = \frac{\int^{2\kfe}_0  \mathcal{C}^2_{\rm en}(q) f_{\kappa/\sigma/\eta}(q)}{\int^{2\kfe}_0 f_{\kappa/\sigma/\eta}(q)}\,,
\label{eq:ceq_av}
\eeq
where 
\beq
f_{\kappa/\sigma/\eta}(q)= \frac{q^3}{(q^2+\qtf^2)^2} \left(1- \frac{q^2}{4 \kfe^2}\right)~I_{\kappa/\sigma/\eta}(q)\,. 
\eeq
From Eq.~\ref{eq:Ins} and the momentum averages defined above we can deduce that the typical momentum transfer $q \approx {\rm few} \times \qtfe < \kfe$, and that in general the typical momenta relevant for the calculation of $\kappa$ are smaller than those relevant for $\sigma$ and $\eta$. $\langle \mathcal{C}_{\rm enp}^2\rangle_{\kappa}$ and $\langle \mathcal{C}_{\rm enp}^2\rangle_{\eta}$ are plotted in Fig.~\ref{fig:cenp_core} for the typical densities expected in the neutron star core. 
\begin{figure}[h]
 \centering
  \includegraphics[width=4in]{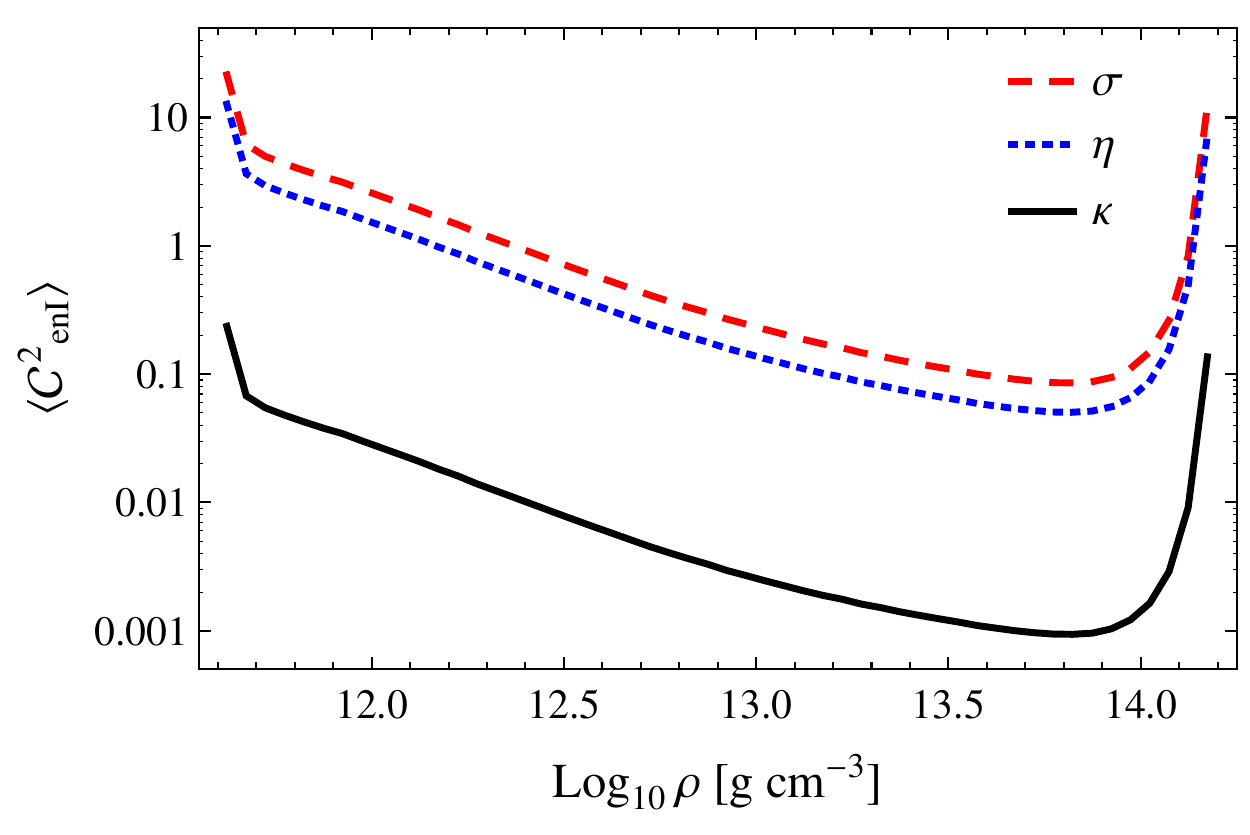}
  \caption{$\langle \mathcal{C}_{\rm enI}^2\rangle_{\kappa/\sigma/\eta}$ in the crust as a function of density using $\vnI$ derived from data presented in Ref.~\cite{kobyakov2013dynamics}}
\label{fig:ceniavg}
\end{figure}
In the crust, the ion density-density correlation function $\chi_I(q)$ has a strong momentum dependence (c.f. Eq.~\ref{eq:chiI2}) , and consequently the stronger momentum dependence of $\cenI(q)$ implies that the differences between the momentum averaged couplings for $\kappa$, $\sigma$, and $\eta$ will be more pronounced than in the core. Using the results for the crust composition and EoS used in \cite{kobyakov2013dynamics} we obtain the density dependence of the momentum averaged effective couplings $\langle {\mathcal C}_{enI} \rangle_{\kappa/\sigma/\eta}$ shown in Fig.~\ref{fig:ceniavg}. 

By using the momentum averaged couplings we can factor $\mathcal{C}^2_{\rm en}(q)$ out of the $q$ integration in Eqs.~\ref{eqn:Lambda_e_kappa}, \ref{eqn:Lambda_e_sigma}, and \ref{eqn:Lambda_e}, and we obtain analytic results for the inverse mean free paths given by 
\bea
\langle \lambda_{en} \rangle_\kappa^{-1} & = &  \frac{4 \pi^2\alfa^2 \langle \mathcal{C}_{\rm en}^2\rangle_\kappa}{5} ~\frac{m_n^2~T^2}{\qtf^3}~\Phi_\kappa\left(\frac{\qtf}{\kfe}\right) \,, \\
\langle \lambda_{en} \rangle_\sigma^{-1} & = &  \frac{\pi^2 \alfa^2 \langle \mathcal{C}_{\rm en}^2\rangle_\sigma}{3}~\frac{m_n^2~T^2}{\kfe^2\qtf}~\Phi_\sigma\left(\frac{\qtf}{\kfe}\right) \,,  \\
\langle \lambda_{en} \rangle_\eta^{-1} & = & \pi^2\alpha^2\langle \mathcal{C}_{\rm en}^2 \rangle_\eta~\frac{m_n^2~T^2}{\kfe^2\qtf}~\Phi_{\eta}\left(\frac{\qtf}{\kfe}\right) \,,
\eea
where 
\bea
\Phi_\kappa(x) & = &  \left(1+\frac{3x^4}{16}\right) \frac{2}{\pi} {\rm ArcTan}\left[ \frac{2}{x} \right] +\frac{x}{\pi} -\frac{3 x^3}{4\pi}  \,, \\
\Phi_\sigma(x) & = & \left(1+\frac{3x^2}{4}\right) \frac{2}{\pi} {\rm ArcTan}\left[ \frac{2}{x} \right] -\frac{3x}{\pi}  \,, \\
\Phi_{\eta}(x) & = & \frac{1}{8\pi}\left(4+x^2\right)\left(4+5x^2\right) {\rm ArcTan}\left[ \frac{2}{x} \right] -\frac{x}{\pi}\left(\frac{13}{3}+\frac{5x^2}{4}\right) \,,
\eea
are normalized so that $\Phi_{\kappa/\sigma/\eta}(0)=1$. 

The corresponding thermal and electrical conductivity, and the shear viscosity are given by the following analytic expressions  
\beq
\kappa_{\rm en}  = \frac{5}{36 \pi^2 \alfa^2 \langle \mathcal{C}_{\rm en}^2\rangle_\kappa}
~\frac{\kfe^2 \qtf^3}{m_n^2~T}~\Phi^{-1}_\kappa\left(\frac{\qtf}{\kfe}\right)  \,,
\label{eq:kappa_e}
\eeq
\beq
\sigma_{\rm en} =  \frac{1}{ \pi^4 \alfa \langle \mathcal{C}_{\rm en}^2\rangle_\sigma}
~\frac{\kfe^4 \qtf}{m_n^2~T^2}~\Phi^{-1}_\sigma\left(\frac{\qtf}{\kfe}\right)\,,
\label{eq:sigma_e}
\eeq
\beq
\eta_{\rm en} = \frac{1}{15\pi^4\alpha^2\langle \mathcal{C}_{\rm en}^2 \rangle_\eta}~\frac{\kfe^6\qtf}{m_n^2~T^2}~\Phi^{-1}_\eta\left(\frac{\qtf}{\kfe}\right)\,,
\label{eq:eta_e}
\eeq
respectively. For the fiducial values $\kfe=100$ MeV, $T=10^8$ K, and $\qtf \approx 30$ MeV, the above formulae predict $\kappa\approx 10^{23}$ erg cm$^{-1}$ s$^{-1}$ K$^{-1}$, $\sigma \approx 10^{29}$ s$^{-1}$ and $\eta\approx 10^{18}$ g cm$^{-1}$ s$^{-1}$ when we set $\langle \mathcal{C}_{\rm en}^2\rangle=1$.    

Now we will consider the case when neutrons are in the superfluid state.  As already mentioned,  neutrons are likely to form s-wave Cooper pairs in the crust, and p-wave Copper pairing is a possibility in the core. While s-wave superfluidity in the crust appears rather robust with critical temperatures for $^1S_0$ pairing in the range $\tcn = 10^8 - 10^{10}$ K, it remains unclear if p-wave pairing occurs in the core. Recent calculations, which account for non-central interactions and polarization effects in the medium, favor the smaller values $\tcn < 10^8 $ K indicating that p-wave pairing is fragile and may be unlikely at typical temperatures encountered in the neutron star core \cite{SchwenkFriman:2004}. Nonetheless, for completeness we entertain the possibility of neutron superfluidity both in the crust and in the core. When $T \ll \tcn$  quasi-particle excitations are suppressed by the factor $\exp{(-\tcn/T)}$ and under these conditions, $S_n(\omega,q)$ is dominated by $\omega = q v_n$ corresponding to interactions with the low energy Goldstone mode associated with the breaking of the $U(1)$ baryon number symmetry in the superfluid ground state.  For a weakly coupled neutron superfluid 
 \begin{equation}
S_n(\omega,q) = \frac{\pi n_n q}{m_n v_n}\left[\frac{\delta(\omega-v_n q)}{1-e^{-\beta \omega}}+\frac{\delta(\omega+v_n q)}{e^{-\beta \omega}-1}\right]~ \,,
\label{eq:s_super}
\end{equation}
where $n_n$ is the neutron density and $v_n \approx \kfn/(\sqrt{3}~m_n)$ is the velocity the superfluid phonon mode \cite{Son2006197}. The relevant frequency integrals for electron mean free paths due to electron collisions with superfluid neutrons are given by  
\bea
 I_{\kappa}(q)&=& \frac{n_n q^2 \beta  (2 \pi ^2+(6 {\kfe}^2-q^2) {v_n}^2 \beta ^2) \text{Csch}(\frac{q v_n \beta }{2})^2}{8 m_n \pi ^2}\,, \\
I_{\sigma}(q)&=& \frac{{n_n} q^2 \beta \ \text{Csch}(\frac{q {v_n} \beta }{2})^2}{4 {m_n}}\,,\\
 I_{\eta}(q) &=& \frac{3n_nq^2\beta(4\kfe^2-q^2)\text{Csch}(\frac{q v_n \beta }{2})^2}{4\kfe m_n} \,.
\eea
Using these expressions the conductivities can be calculated numerically.  
\section{Results and Discussion} 
\label{sec:results}
\subsection{Electron Transport in the Core}
We find that electron-neutron scattering is most relevant in the core when neutrons are in the normal phase and protons are superconducting.  This is because proton superconductivity suppresses electron scattering from the other electrons, muons, and protons. When the protons are in the normal phase, electrons interact mainly through the current-current interaction because in this case the interaction is only weakly screened by dynamical effects due to the Landau damping of transverse plasmons \cite{Heiselberg:1993cr}. In the superconducting state, the transverse plasmon is massive due to the Meissner effect and the inverse screening length (proportional to $\Delta_{\rm p}$) suppresses the electron-electron, electron-muon, and electron-proton scattering. Electron-proton scattering is additionally suppressed by the factor $\simeq \exp{(-\tcp/T)}$ due to the gap in the proton particle-hole spectrum.  In what follows we discuss the relevance of electron-neutron scattering in the core for typical conditions and choose two fiducial values of the proton critical temperature $\tcp=10^9$ K and $\tcp=10^{10}$ K, which are assumed to be independent of density.  

We begin by discussing the relevance of electron-neutron scattering for the electron thermal conductivity. In the core electrons can scatter off other electrons, muons, protons, and neutrons. Hence the full electron thermal conductivity is given by 
\begin{equation}
\kappa_{\rm e} = \left(\frac{1}{\kappa_{\rm ref}}+\frac{1}{\kappa_{ en}}\right)^{-1} \,, 
\end{equation} 
where 
\begin{equation} 
\kappa_{\rm ref} = \left(\frac{1}{\kappa_{ e e}}+\frac{1}{\kappa_{ e \mu}}+ \frac{1}{\kappa_{ e p}}\right)^{-1} \,, 
\end{equation} 
is the contribution to the electron thermal conductivity due to electron-electron, electron-muon, and electron-proton scattering considered in earlier work.  When $T \ll \tcp$, Shternin and Yakovlev find that 
\beq 
\kappa_{\rm ref }(T\ll \tcp)= \frac{5}{24 \alpha}~\kfe^2~\frac{\Delta_p}{T}~f \,,
\label{eq:kappa_SY07}
\eeq
where the factor $f=\kfp^2/(\kfe^2+\kfmu^2)\approx 1$ includes the correction in regions where the muon fraction is not negligible \cite{Shternin:2007}. Comparing Eq.~\ref{eq:kappa_SY07} to the result we obtained for electron-neutron scattering in Eq.~\ref{eq:kappa_e} we can estimate when the latter will dominate. To obtain a simple expression we neglect $\kfmu$, and set $\kfp=\kfe$ to find that when 
\beq
 \langle \mathcal{C}_{\rm en}^2\rangle_{\kappa} \gtrsim \frac{16\sqrt{\alpha}}{3\pi^{7/2}} \sqrt{\frac{\kfe}{m_n}}~\frac{\kfe}{\Delta_p}\approx 0.27 ~\left(\frac{\kfe}{100 ~{\rm MeV}} \right)^{3/2}~\left(\frac{1~{\rm MeV}}{\Delta_p}\right)\,,
\eeq
electron-neutron scattering dominates. 

\begin{figure}[h] 
   \centering
   \includegraphics[width=3.56in]{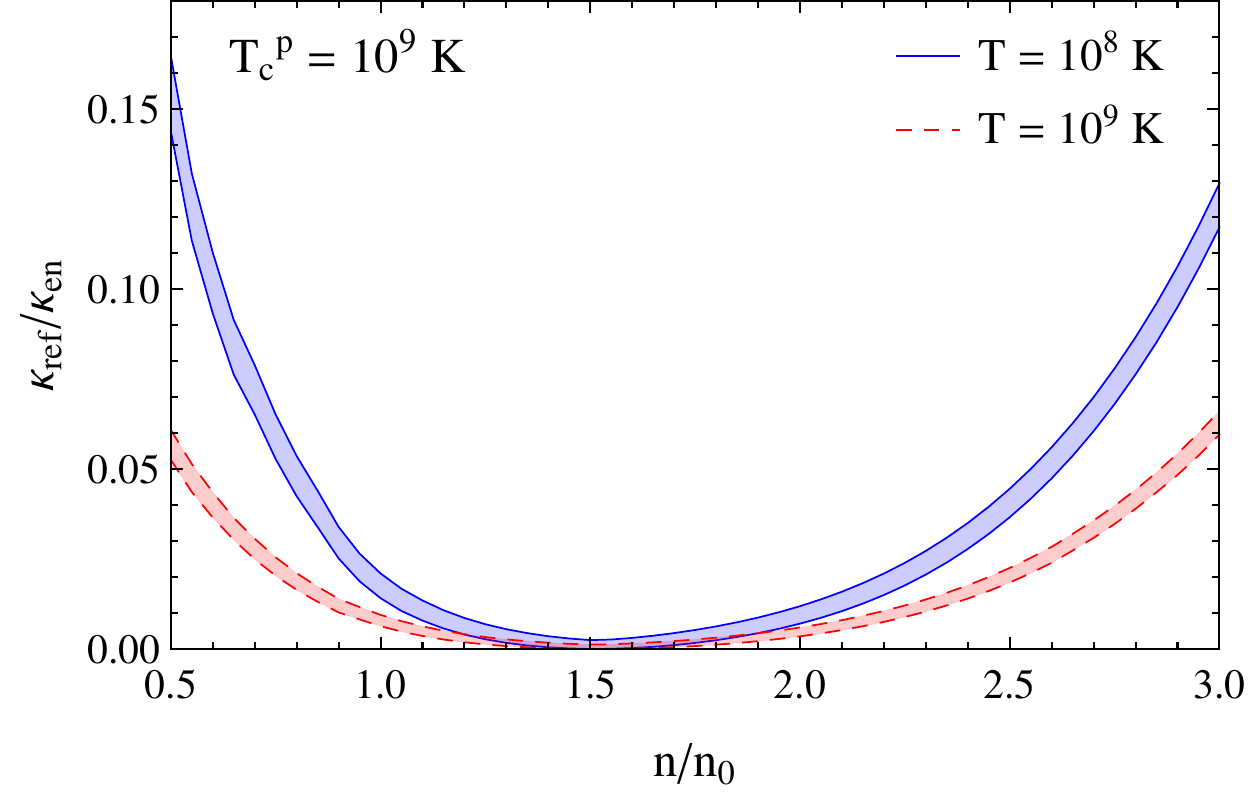} 
   \includegraphics[width=3.5in]{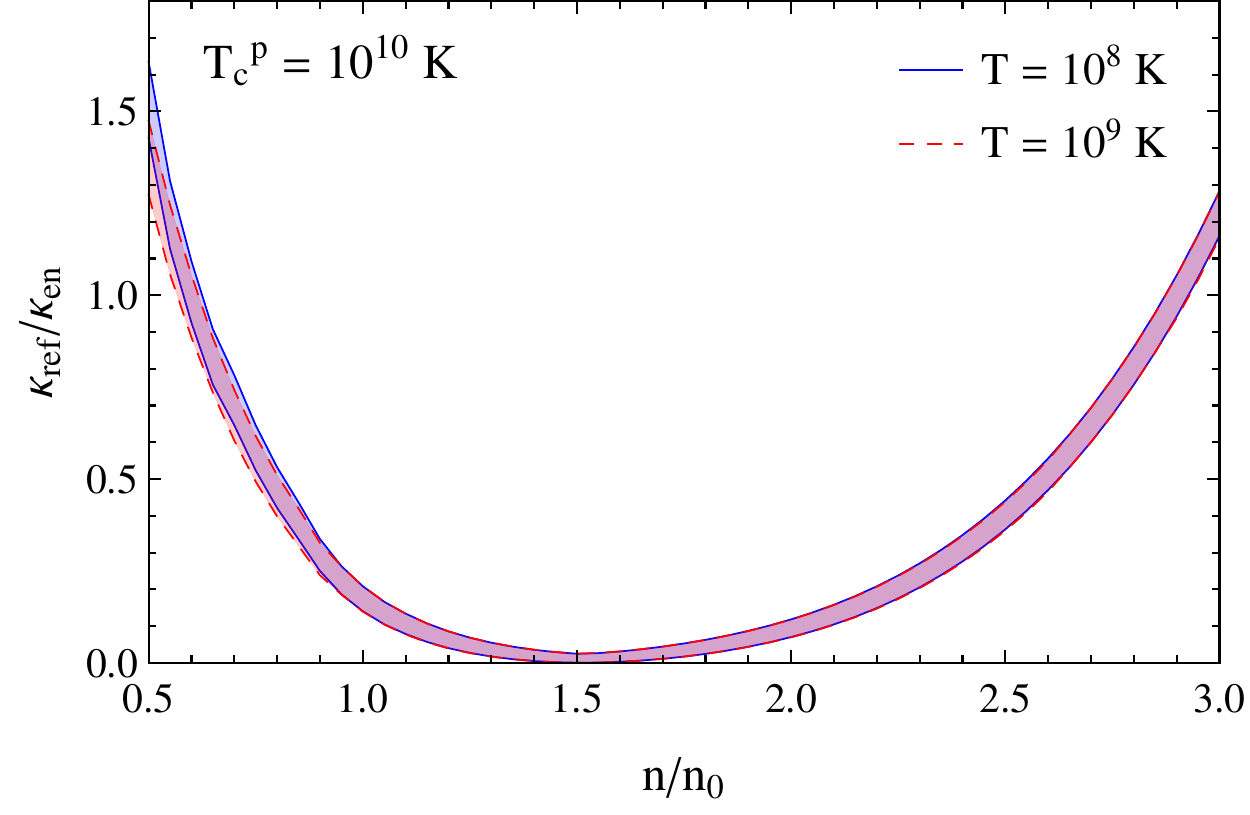} 
   \caption{The ratio $\kappa_{\rm ref}/\kappa_{en}$ for two proton critical temperatures, $T_c^p=10^9$ K (left) and $T_c^p = 10^{10}$ K (right) for densities of relevance to the core. When the $\kappa_{\rm ref}/\kappa_{en} > 1$, electron-neutron scattering dominates. }
   \label{fig:kappa_core}
\end{figure}

In Fig.~\ref{fig:kappa_core} we show the ratio $\kappa_{\rm ref}/\kappa_{en}$ to asses the relative importance of electron-neutron scattering.  $\kappa_{\rm ref}$ is calculated using the fitting formula from \cite{Shternin:2007} and is described in Appendix \ref{app:previous_kappa} for reference.  When  $\kappa_{\rm ref}/\kappa_{en}>1$ electron-neutron scattering is the dominant scattering mechanism and  from the figure we can deduce that electron-neutron scattering is unlikely to be important when $\tcp$ is small and $T \gtrsim \tcp$. This is because a smaller gap $\Delta_p \simeq 1.76~ \tcp$ results in weaker screening of electron-electron and electron-muon scattering, and when $T\gtrsim \tcp$, electron-proton scattering becomes relevant and further reduces the electron mean free path. For the larger $\tcp \simeq 10^{10}$ K, results shown in the right panel of Fig.~\ref{fig:kappa_core} indicate that electron-neutron scattering can be relevant both at low density near the crust-core boundary, and at higher density deep inside the core. Here, because $T > \tcp$ both $\kappa_{\rm ref}$ and $\kappa_{en}$  scale as $1/T$. Thus, their ratio is independent of temperature and the bands shown in the right panel overlap. 

Similarly, the total electron shear viscosity in the core  is given by 
\begin{equation}
\eta_{\rm e} = \left(\frac{1}{\eta_{\rm ref}}+\frac{1}{\eta_{ en}}\right)^{-1} \,,
\end{equation} 
where 
\begin{equation} 
\eta_{\rm ref} = \left(\frac{1}{\eta_{ e e}}+\frac{1}{\eta_{ e \mu}}+ \frac{1}{\eta_{ e p}}\right)^{-1} \,, 
\end{equation} 
is the contribution to the shear viscosity due to electron scattering off other electrons, protons, and muons considered in earlier work. For $T \ll \tcp$, the case of interest to us here, Shternin and Yakovlev find that 
\beq 
\eta_{\rm ref}(T\ll \tcp)= \frac{\xi}{9\pi^4 \alpha^{5/3}}~\frac{\kfe^5}{T^{2}}~\left(\frac{\Delta_p}{\kfp}\right)^{1/3}~f',
\label{eq:eta_SY08}
\eeq
where $\xi \approx 1.7$, and the factor $f'=\kfp \kfe/(\kfe^2+\kfmu^2)\approx 1$ includes the correction in regions where the muon fraction is not negligible \cite{Shternin:2008}. Comparing Eq.~\ref{eq:eta_SY08} to the result we obtained for electron-neutron scattering in Eq.~\ref{eq:eta_e} we can estimate that electron-neutron scattering will dominate when the induced coupling 
\beq
 \langle \mathcal{C}_{\rm en}^2\rangle_{\eta} \gtrsim \frac{6\alpha^{1/6}}{5\sqrt{\pi}\xi} \left(\frac{\kfe}{m_n}\right)^{3/2}\left(\frac{\kfe}{\Delta}\right)^{1/3} \approx  0.03 ~\left(\frac{\kfe}{100 ~{\rm MeV}} \right)^{11/6}~\left(\frac{1~{\rm MeV}}{ \Delta_p}\right)^{1/3}\,.
\eeq
As before, in deriving the above criterion, we have neglected $\kfmu$ and set $\kfp= \kfe$. 
\begin{figure}[h] 
   \centering
   \includegraphics[width=3.65in]{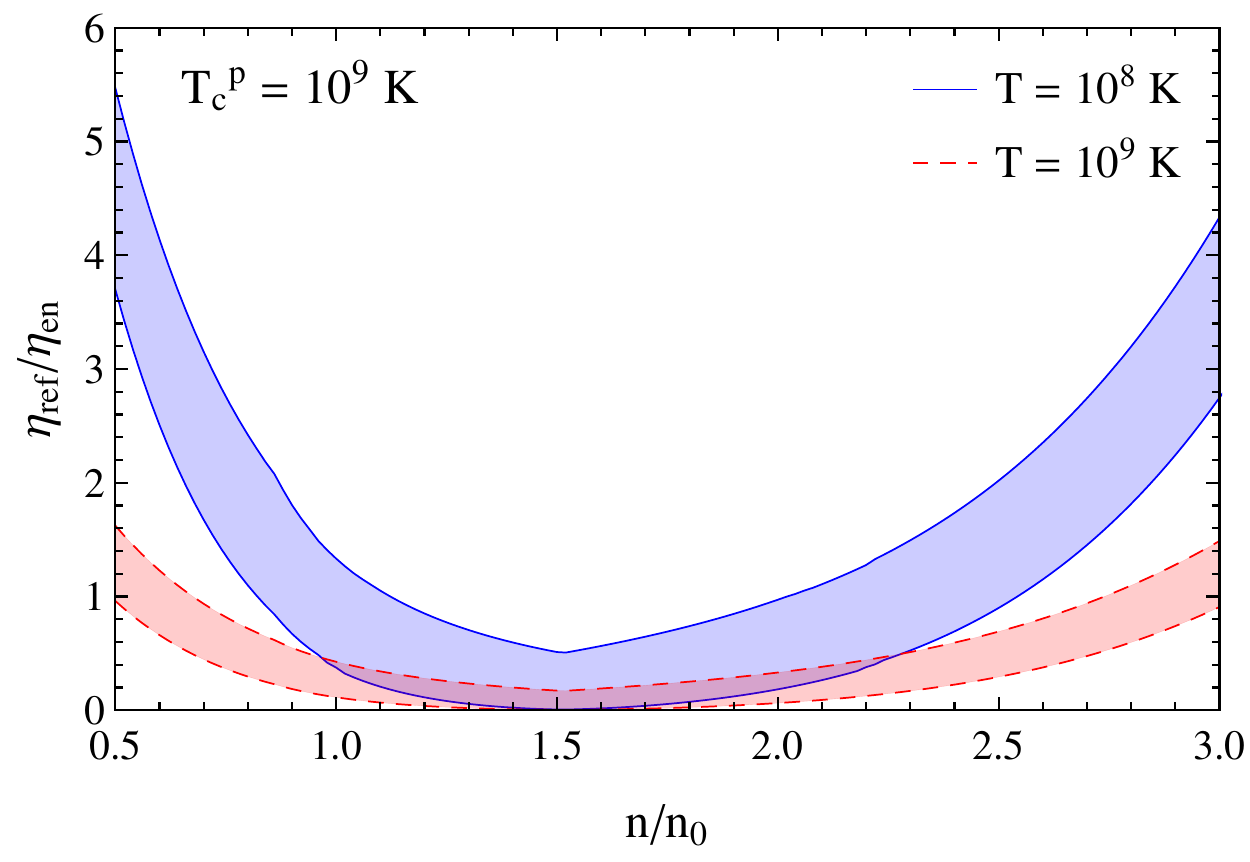} 
   \includegraphics[width=3.7in]{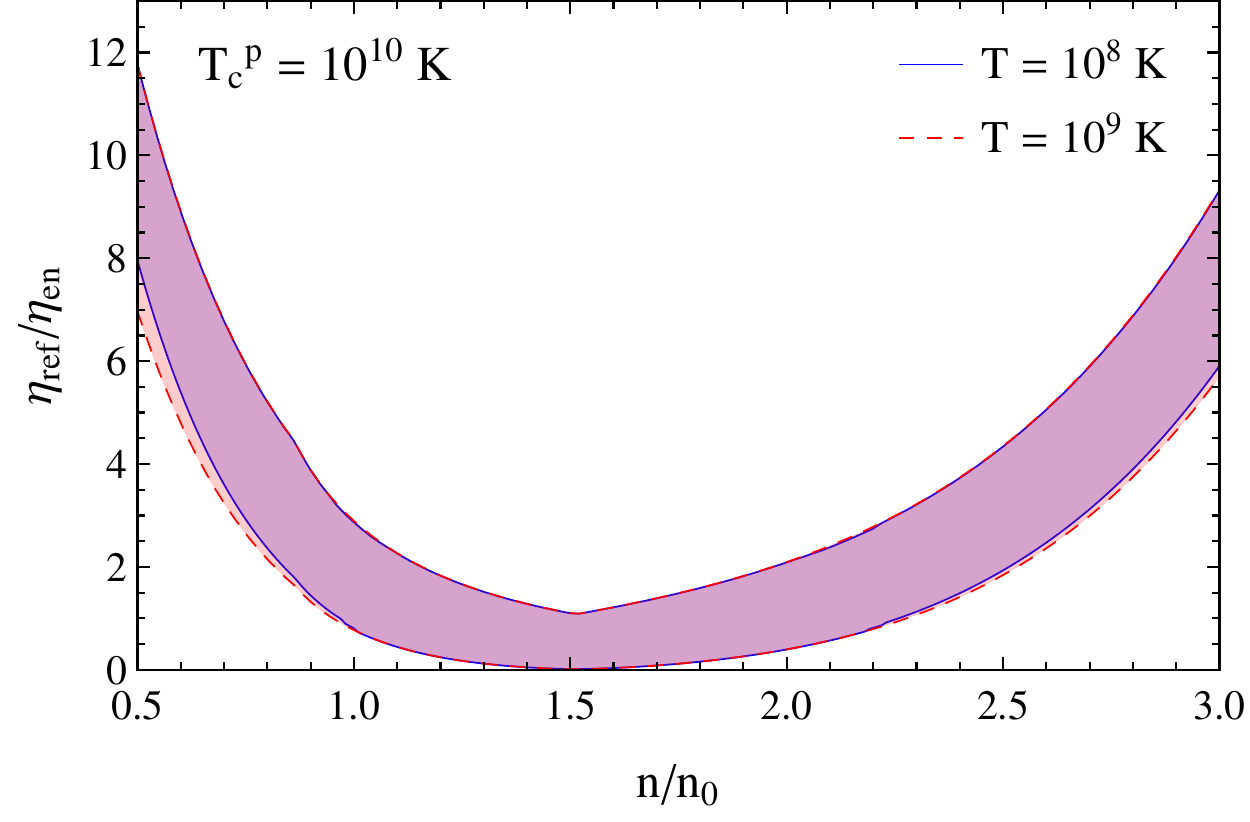} 
   \caption{The ratio $\eta_{\rm ref}/\eta_{en}$ for two proton critical temperatures, $T_c^p=10^9$ K (left) and $T_c^p = 10^{10}$ K (right) for densities in the core.}
   \label{fig:eta_core}
\end{figure}

In Fig.~\ref{fig:eta_core} we show the ratio $\eta_{\rm ref}/\eta_{en}$ for two values of fiducial temperature in the core $T=10^8$ and $T=10^9$ K, and for two representative values of the proton critical temperature to asses the relative importance of electron-neutron scattering.  $\eta_{\rm ref}$ is calculated using the fitting formula from \cite{Shternin:2008} and is described in Appendix \ref{app:previous_eta} for reference. As before, in regions where $\eta_{\rm ref}/\eta_{en}>1$, electron-neutron  scattering is the dominant scattering mechanism for electrons.  The results in the figure indicate that electron-neutron scattering is relevant everywhere in the core. Further, because $\eta_{\rm ref}$ is a weaker function of the superconducting gap compared to $\kappa_{\rm ref}$, we find that $\eta_{\rm ref}/\eta_{en} >1$ even when $T\simeq \tcp$.

The general trend that the electron-neutron contribution is more relevant for $\eta$ rather than $\kappa$, and that it remains relevant even when proton superconductivity is weak or absent, can be understood by noting that  screening is more important for $\kappa$ than it is for $\eta$. This is because low momentum scattering with energy transfer $\omega \simeq T$ can make an important contribution to $\kappa$ and is reflected by the fact that $\kappa \propto \Lambda^3$ where $\Lambda$ is the momentum scale set by the physics of screening, while $\eta \propto \Lambda$ (c.f. the dependence on $\qtf$ in Eqs. \ref{eq:kappa_e} and \ref{eq:eta_e}). In the case of the density-density interaction that we have considered between electrons and neutrons $\Lambda=\qtf \approx (4\alpha m_p \kfp/\pi)^{1/2}$, while for the current-current interaction between electrons considered in \cite{Shternin:2007,Shternin:2008} the relevant scale of the screening momentum is $\Lambda \approx (\pi \alpha \kfp^2 \Delta_p)^{1/3}$ when protons are superconducting, and $\Lambda \approx (2 \alpha T \kfp^2)^{1/3}$ when protons are normal.    

We have calculated both $\kappa_{en}$ and $\eta_{en}$ for the case when neutrons are superfluid and found them to be too large compared to $\kappa_{\rm ref}$ and $\eta_{\rm ref}$ to be relevant. Here, electron scattering occurs either by absorption or emission of the superfluid phonon with energy $\omega = q v_n$. Since large energy transfer is exponentially  suppressed due to degeneracy and typical $\omega \simeq T$, the electron-phonon scattering is highly peaked in the forward direction and contributes little to the electron transport properties.  Finally, we note that the electrical conductivity is only relevant when protons are normal, and in this case we find that electron-neutron scattering can be as relevant as  electron-proton scattering when $\langle \mathcal{C}^2_{enp} \rangle \gtrsim 1$. However, this warrants a careful study of the induced interaction between the electron and neutron currents mediated by transverse plasmons in the normal state and is beyond the scope of this study.


\subsection{Electron Transport in the Crust}
 
In the inner crust, ions form a crystal and electron-ion scattering is suppressed due to correlations for $T<T_P$ where $T_P=\hbar \omega_P/k_B$ is the ion plasma temperature and $\omega_P= \sqrt{4\pi \alfa Z^2 n_I/M_I} $ is the plasma frequency of ions with charge $Z$, mass $M_I$, and number density $n_I$. The dominant electron scattering processes considered in earlier work were due to electron-phonon and electron-impurity interactions. When the impurity concentration is negligible, the electron contribution to $\kappa$, $\sigma$, and $\eta$ at low temperature is limited by the emission or absorption of lattice phonons by electrons and has been studied in earlier work  \cite{FlowersItoh:1976,Yakovlev:1980,Itoh:1984,Baiko:1996}. The importance of Umklapp scattering was realized early in Ref.~\cite{FlowersItoh:1976} because this allows the electron momentum to change by a large amount, ${\bf K}=(2\pi/a) (n_x {\bf \hat{x}}  +n_y {\bf\hat{y}} + n_z {\bf\hat{z}})$, where $n_i$ are integers, even for relatively small energy transfer $\omega \simeq T \ll |{\bf K}|$ . For this reason, electron-ion scattering remains very effective down to low temperatures until the Umklapp processes are frozen out by the small band gap in the electron spectrum. This occurs when $T < T_{\rm um}$ where $T_{\rm um} \approx  (\alpha/9\pi)Z^{1/3} T_P$ is called the Umklapp temperature. In the inner crust where $Z\approx 40$,  $T_{\rm um} \approx 10^{-3} T_P$.  

By comparing our results in the inner crust to those obtained from only considering electron-phonon Umklapp scattering from Ref.~\cite{Baiko:1996}, we find that our values for $\kappa$ and $\sigma$ are much too large (typically by a factor of about 100 or more) to be relevant for temperatures in the range $10^7-10^9$ K. A similar comparison between our results for $\eta$ with those presented in Ref.~\cite{Chugunov} shows that electron-neutron scattering is also too weak to be relevant in this case. At very low temperature when $T \lesssim10^6$ K the Umklapp process is frozen out, and in this case electron-neutron scattering can become relevant if neutrons are in the normal phase and the impurity concentration can be neglected. However, since neither of these conditions seem likely in the crust we do not present a detailed comparison. 

\section{Conclusions and caveats}\label{sec:conclusions}
We have identified a new mechanism for electron scattering off neutrons induced by protons in the core and ions in the crust.  Using simple models of the neutron-proton interaction in the core and the neutron-ion interaction in the crust, we have estimated the strength of this induced interaction. In calculating the electron thermal conductivity, shear viscosity, and electrical conductivity, the coupling was characterized by  a density-dependent, dimensionless parameter $\langle \cenp^2\rangle$ in the core, and $\langle \cenI^2\rangle$ in the crust. Our main findings are: 

\begin{itemize}
\item When protons are superconducting with $T_c^p \gtrsim 10^9$ K, electron-neutron scattering is a relevant process that limits the electron mean free paths and determines the  electron thermal conductivity and shear viscosity in the neutron star core. Our calculations indicate the shear viscosity in the vicinity of the crust-core transition can be  $5-10$ times larger than earlier estimates.  This enhancement may be relevant in the context of damping hydrodynamic modes and r-modes, and could have implications for neutron star spin evolution and gravitational wave instabilities (e.g. see \cite{cutler,Andersson:1997xt}). In contrast, changes to the thermal conductivity are relatively small and given current uncertainties in their calculation these changes are unlikely to be of much interest for neutron star phenomenology.   

\item In the crust, electron-neutron scattering is unlikely to be relevant because (i) electron Umklapp scattering off the ion lattice is efficient for typical temperatures of interest, (ii) at low temperature when Umklapp scattering is supressed, electron-impurity scattering will likely dominate, and (iii)  neutrons are likely to be in the superfluid state in the crust and in this case electron-neutron scattering is highly suppressed.  
 \end{itemize}

The results we presented are sensitive to the momentum dependent electron-neutron couplings  $\cenp(q)$ and $\cenI(q)$. Although these coupling were well determined at low momentum because they were directly related to the EoS, the finite momentum component warrants further study.  Calculations of the Fermi liquid parameters and effective interactions  in asymmetric matter are being pursued using realistic nucleon-nucleon potentials and will be reported in future work. 

In addition, here we have only considered the electron coupling to neutron density fluctuations and have ignored the current-current coupling. We argued that the current-current interaction would be subdominant because protons, and to a lesser extent neutrons, can be treated as non-relativistic particles. However, when protons in the core are in the normal phase, the current-current interaction is stronger because it is only weakly screened by Landau damping. This may well compensate for the smaller proton velocity and warrants further investigation, especially to ascertain if electron-neutron scattering could be relevant in the absence of proton superconductivity in the core. We note that the formalism to incorporate the electron coupling to the density and current of a multi-component interacting medium exists and has been used in the context of studying neutrino scattering in hot and dense matter in Ref.~\cite{Horowitz:1990it,Reddy:1998hb}. This formalism can be adapted to study cold matter with pairing correlations and will be explored in future work. 

Finally, we note that in the crust, Bragg scattering of unbound neutrons from the ion lattice will result in a distorted Fermi surface and band structure for the neutrons. This implies that Umklapp processes involving neutrons can become important. Since such processes enable large momentum transfer they warrant further study before we can discount the relevance of electron-neutron processes in the crust. 

\begin{acknowledgments}
We thank Charles Horowitz, David Kaplan, Chris Pethick, Martin Savage and Dima Yakovlev for useful discussions and Andrew Steiner for reading the manuscript.  The work of S. R. and B. B. was supported by the DOE Grant No. DE-FG02-00ER41132 and by the Topical Collaboration to study {\it Neutrinos and nucleosynthesis in hot and dense matter}. The work of S. R.  and E. R. was also supported by the NUCLEI SciDAC program.       
\end{acknowledgments}


\appendix

\section{A derivation of the electron-neutron induced interaction}
\label{app:encore}

To illustrate how the induced interaction arises we consider the case of electron scattering from an interacting liquid of neutrons and protons at zero temperature. For simplicity, we shall assume that protons and neutrons are non-relativistic, and only consider the Coulomb coupling of electrons to the proton density. The linear response formalism to describe scattering off the density fluctuations in a liquid in terms of the density-density correlation function $\Pi^p(\omega,q)$ is derived and discussed in Ref.~\cite{FetterWalecka:2003}. Explicitly, the differential cross-section per unit volume for an electron with momentum $k$ to scatter from density fluctuations in a proton liquid is given by 
\beq
\frac{1}{V}~\frac{d \sigma}{d\omega dq}(k) = -\frac{q}{2 \pi^2}\left (1-\frac{q^2}{4k^2} \right)~\mathcal{U}^2 _{\rm ep}(q)~ \im~ {\Pi^p}(\omega,q); \quad \mathcal{U} _{\rm ep}(q)
= \frac{-4\pi \alpha}{q^2+\qtf^2} \,,
\label{eqn:dsigma_ep}
\eeq
where $\omega$ and $q$ are the energy and momentum transfer from the electron to the medium, $\alpha=1/137$ is the fine structure constant, and $\mathcal{U} _{\rm ep}(q)$ is the effective interaction in the medium between electrons and protons which includes the effects of screening in the plasma through the Thomas-Fermi screening momentum, $\qtf$.  

A generalization of this formalism to a two component liquid is outlined in \cite{Horowitz:1990it,Reddy:1998hb} where it was used to study neutrino scattering off neutrons and protons. Using this generalized formalism to describe scattering of relativistic electrons off neutron and proton density fluctuations, we find that the differential cross-section per unit volume for an electron with momentum $k$ to scatter from a neutron-proton liquid is given by
\beq
\frac{1}{V}~\frac{d \sigma}{d\omega dq}(k) = -\frac{q}{2 \pi^2}\left (1-\frac{q^2}{4k^2} \right)\frac{1}{(q^2+\qtf^2)^2}~
(\begin{matrix} 0, & 4 \pi \alpha\\  \end{matrix} )~ \im~ {\bf \Pi}(\omega,q)~\left( \begin{matrix} 
      0 \\ 
     4 \pi \alpha \\
   \end{matrix} \right)\,,
\label{eqn:dsigma}
\eeq
where ${\bf \Pi}(\omega,q)$ is the density-density correlation function of the two component liquid which we shall define explicitly below. The coupling between the neutron and proton components of the liquid is incorporated by summing a class of particle-hole diagrams most relevant at long-wavelengths within the Random Phase Approximation (RPA) \cite{FetterWalecka:2003}.  In this approximation (which satisfies current conservation), a closed form expression for the time-ordered correlation function ${\bf \Pi}(\omega,q)$ exists and is given by 
 \bea
 {\bf \Pi}_{\rm RPA}(\omega,q)&= &({\bf 1}- {\bf V}~{\bf \Pi^0})^{-1}~{\bf \Pi^0}~\,, \\
   {\bf \Pi^0}&=&
   \left( \begin{matrix} 
      {\bf \Pi}^0_n & 0 \\
      0 & {\bf \Pi}^0_p \\
   \end{matrix} \right)\,,\quad
   {\bf V}=
   \left( \begin{matrix} 
      V_{nn} & V_{np} \\
      V_{np} & V_{pp} \\
   \end{matrix} \right)\,,
\label{eqn:pi_rpa}
\eea  
where ${\bf V}$ is the interaction matrix that describes the strong interactions between quasi-particles in the liquid.  The neutron and proton time-ordered density-density correlation functions ${\bf \Pi}^0_n$ and ${\bf \Pi}^0_p$ are approximated by the expressions obtained for a non-interacting Fermi gas  
\beq
\Pi^0_j(\omega,q)= -2i~\int \frac{dk_0~d^3k}{(2\pi)^4}~\G_j (k_0+\omega,|\vec{k}+\vec{q}|) \G_j(k_0,k) \,,
\label{eqn:pi_0}
\eeq 
where $\G_j(k_0,k)$ is the single particle Greens function \cite{FetterWalecka:2003}.  

To make explicit the real and imaginary parts, the correlation function can be written as 
\beq
{\bf \Pi}^0_j = \chi_j + i \Phi_j \,.
\eeq
Of particular interest to our study here is the case when $ \Phi_p \approx 0  \ll |\chi_p|$. This is realized when protons are superconducting and the energy transfer $\omega $ is small compared to the gap in the excitation spectrum, $\Delta_p$. Further, when $|V_{nn} \chi_n| \ll 1$ and $|V_{pp} \chi_p| \ll 1$, we find that in the limit $\Phi_p \rightarrow 0$ the differential cross-section per unit volume is given by 
\beq 
\frac{1}{V} \frac{d\sigma}{ d\omega dq} = -\frac{8 \alpha^2 q}{(q^2+q_{TF}^2)^2}\left(1-\frac{q^2}{4 k_{Fe}^2}\right) V^2_{pn}\chi^2_{p}~\Phi_n\,. 
\label{eqn:dsigma_rpa}
\eeq
Here, electrons excite neutron particle-hole states indirectly through the proton polarization. Comparing Eq.~\ref{eqn:dsigma_rpa} with Eq.~\ref{eqn:dsigma_ep} we define the induced electron-neutron interaction 
\beq
\mathcal{U} _{\rm enp}(\omega,q)
=\frac{-4\pi \alfa~\cenp(\omega,q) }{q^2+\qtf^2} \,,
\label{eq:epn}
\eeq 
where $\cenp(\omega,q)=V_{np}\chi_p(\omega,q)$.


\section{Comparison to Previous Results}\label{app:previous}

In Fig.~\ref{fig:kappa_core} and Fig.~\ref{fig:eta_core} we have used results for $\kappa_{\rm ref}$ and $\eta_{\rm ref}$ from previous calculations where electron-electron, electron-muon, and electron-proton scattering was considered.  Here we write down explicitly the formulae that were used. 

\subsection{Thermal Conductivity}
\label{app:previous_kappa}
From  \cite{Shternin:2007} we find that the electron contribution to the thermal conductivity due to electron-electron, electron-muon and electron-proton scattering is given (in natural units) by
\begin{equation}
\kappa_{ei} = \frac{\pi^2 T n_e }{3 m_e^* \nu_{ei}^{\kappa}} \,,
\label{eq:kappa_ei}
\end{equation}
where $n_e$ is the number density of electrons, $T$ is the temperature, $m_e^* = \mu_e \approx k_{Fe}$, and $\nu_{ei}^{\kappa}$ is the frequency of electron-i collisions where i stands for electrons, muons, or protons. The collision frequency is dominated by the exchange of transverse plasmons because scattering due to the exchange of longitudinal plasmons is strongly screened in the static limit.  In Eq.~\ref{eq:kappa_ei} the coupling between the heat transport of electrons and muons is neglected and was shown to be unimportant for strongly degenerate conditions in neutron star cores \cite{Shternin:2007}.

In the absence of proton superconductivity, the collision frequency for thermal conductivity is given by
\begin{equation}
\nu_{ei}^{\kappa} = \frac{24\zeta (3)}{\pi^3}\frac{\alpha^2 T}{m_e^{*}}\frac{k_{\rm Fi}^2 k_{\rm Fe}}{q_{\rm t}^2} \,,
\end{equation}
where $\zeta (z)$ is the Riemann zeta function, $k_{\rm Fi}$ is the Fermi momentum of particle $i$, and $q_{\rm t}$ is the transverse screening momentum, 
\begin{equation}
q_{\rm t}^2 = \frac{4\alpha}{\pi}\sum_i k_{Fi}^2 \,.
\label{eq:qt}
\end{equation}

The total electron thermal conductivity is then given by
\begin{equation}
\kappa_{\rm ref} = \left(\frac{1}{\kappa_{ e e}}+\frac{1}{\kappa_{ e \mu}}+ \frac{1}{\kappa_{ e p}}\right)^{-1} = \frac{\pi^2}{54 \zeta(3)}\frac{k_{\rm Fe}^2}{\alpha} \,,
\end{equation}
only depending on $k_{\rm Fe}$ and not on the temperature---an interesting dependence that is due to the exchange of dynamically screened transverse plasmons.  It is in contrast to our result of $\kappa \propto T^{-1}$ which is due to the exchange of longitudinal plasmons that remain massive in the static limit.

In the presence of proton superconductivity, the transverse plasmons become massive due to the Meissner effect and this restores the $T^{-1}$ behavior of $\kappa_{\rm ref}$. In addition, the number of proton quasiparticles becomes exponentially suppressed.  In this case, from  \cite{Shternin:2007} we find that the thermal conductivity
\begin{equation} \label{eqn:kappaSY}
\kappa_{\rm ref} \rightarrow \kappa_{\rm ref}\frac{1}{R_{\rm tot}^{\kappa}(y,r)} \,,
\end{equation}
where 
\begin{align}
& R_{\rm tot}^{\kappa}(y,r) = p_1 e^{-0.14 y^2} + \frac{1-p_1}{\sqrt{1+p_3 y^2}}\,,  \quad p_1 = 0.48 - 0.17r \,,\quad  p_3 = \left[ (1-p_1)\frac{45 \zeta (3)}{4\pi^2 r} \right]^2 \,,\\
& y = \sqrt{1-t}\left(1.456 - \frac{0.157}{\sqrt{t}} + \frac{1.764}{t} \right) \,, \quad r =  \frac{k_{\rm Fe}^2 + k_{\rm F\mu}^2}{k_{\rm Fp}^2} \,, \label{r} 
\end{align}
and $t=T/\tcp$. Eq.~\ref{eqn:kappaSY} was used to make the plots shown in Fig.~\ref{fig:kappa_core}. For the case of strong proton superconductivity when $y \gg 1$, the thermal conductivity simples to 
\begin{equation}
\kappa_{\rm ref} (T \ll T_c^p) \approx \frac{5}{24\alpha}\frac{k_{\rm Fe}^2k_{\rm Fp}^2}{k_{\rm Fe}^2+k_{\rm F\mu}^2}\frac{\Delta_p}{T} \,.
\end{equation}

\subsection{Shear Viscosity}
\label{app:previous_eta}
For the electron contribution to the shear viscosity we used the analytic results of \cite{Shternin:2008} in the same way as  we used \cite{Shternin:2007} for the thermal conductivity discussed above. The shear viscosity is given by  
\begin{equation}
\eta_{ei} = \frac{n_ek_{\rm Fe}^2}{5 m_e^* \nu_{ei}^{\eta}} \,,
\end{equation}
when the longitudinal plasmon exchange and the coupling between the electron and muon shear viscosity can be neglected.  This is a good approximation here, though it is not as good as in the thermal conductivity case.  In the absence of proton superconductivity, the collision frequencies for shear viscosity are given by
\begin{equation}
\nu_{ei}^{\eta} = \frac{\chi \alpha \pi}{4} \frac{q_{\rm t}^{4/3}}{k_{Fe} m_e^*} T^{5/3} \,,
\end{equation}
where $\chi \approx 6.93$ and $q_{\rm t}$ is given by Eq. \ref{eq:qt}.  Note that this expression (and subsequent expressions) in \cite{Shternin:2008} is missing the factor of $\alpha$. 

The total electron shear viscosity is then given by
\begin{equation}
\eta_{\rm ref} = \left(\frac{1}{\eta_{ e e}}+\frac{1}{\eta_{ e \mu}}+ \frac{1}{\eta_{ e p}}\right)^{-1} = \frac{4}{15\chi\alpha}\frac{k_{\rm Fe}^6}{\pi^3}\left(\frac{1}{q_{\rm t}^4 T^5}\right)^{1/3}  \,,
\end{equation}
where the temperature dependence is due to the exchange of dynamically screened transverse plasmons and is in contrast to our result of $\eta \propto T^{-2}$ due to the exchange of massive longitudinal plasmons.

In the presence of proton superconductivity, this screening of transverse plasmons is no longer dynamical and the transverse plasmons acquire a mass (which restores the $T^{-2}$ behavior of $\eta_{\rm ref}$), and the number of proton quasiparticles becomes exponentially suppressed.  Allowing for proton superconductivity, the shear viscosity becomes
\begin{equation} \label{eqn:etaSY}
\eta_{\rm ref} \rightarrow \eta_{\rm ref}\frac{1}{R_{\rm tot}^{\eta}(y,r)} \,,
\end{equation}
where 
\begin{align}
& R_{\rm tot}^{\eta} = \frac{1-g_1}{(1+g_3 y^3)^{1/9}} + (g_1+g_2)\text{Exp}\left[0.145-\sqrt{0.145^2+y^2}\right] \,, \\
& g_1 = 0.87 - 0.314r \,, ~ g_2 = (0.423 + 0.003r)y^{1/3} + 0.0146y^2 - 0.598y^{1/3}e^{-y} \,, \\
& g_3 = 251r^{-9}(r+1)^6(1-g_1)^9 \,,
\end{align}
and $y$, $t$, and $r$ were defined earlier in appendix \ref{app:previous_kappa}. Although we use Eq.~\ref{eqn:etaSY} to make the plots shown in Fig.~\ref{fig:eta_core} we note that when proton superconductivity is strong $(y \gg 1)$, the shear viscosity is given by the simpler expression
\begin{equation}
\eta_{\rm ref} (T \ll T_c^p) \approx \frac{\xi}{9\pi^4\alpha^{5/3}}\frac{k_{\rm Fe}^6}{T^2}\frac{k_{\rm Fp}^{2/3}}{k_{\rm Fe}^2+k_{\rm F\mu}^2} \Delta_p^{1/3} \,,
\end{equation}
where $\xi \approx 1.71$.

\bibliographystyle{apsrev}
\bibliography{crust_conduction}

\end{document}